\documentclass[12pt,letterpaper]{article}
\usepackage[a4paper, total={7in, 10in}]{geometry}

\usepackage{graphicx}
\usepackage{helvet}
\usepackage{authblk}
\usepackage{hyperref}
\usepackage{amsmath}
\usepackage{amssymb}
\usepackage[utf8]{inputenc}
\usepackage[super,comma,sort&compress]{natbib}
\usepackage{orcidlink}
\usepackage{amsthm}
\newtheorem{assumption}{Assumption}
\newtheorem{theorem}{Theorem}
\usepackage{booktabs}

\makeatletter
\renewcommand{\maketitle}{\bgroup\setlength{\parindent}{0pt}
\begin{flushleft}
  \textbf{\@title}\\[10pt]
  \@author
\end{flushleft}\egroup}
\makeatother

\title{Nationally Consistent, Locally Incomplete: A Bayesian Remote-Sensing Audit of Rooftop Photovoltaic Registries}

\author[1,2,4,*]{Gabriel Kasmi\,\orcidlink{0000-0002-7774-4302}}
\author[1]{Yves-Marie Saint-Drenan\,\orcidlink{0000-0003-1471-1092}}
\author[2,3]{Laurent Dubus\,\orcidlink{0000-0002-3987-646X}}
\author[1]{Philippe Blanc\,\orcidlink{0000-0002-6345-0004}}
\affil[1]{Centre Observation Impacts Energie (O.I.E.), MINES Paris, Université PSL, Sophia-Antipolis, France}
\affil[2]{Réseau de Transport d'Electricité (RTE), Paris La Défense, France}
\affil[3]{World Energy \& Meteorology Council (WEMC), Norwich, UK}
\affil[4]{Lead contact}
\affil[*]{Correspondence: gabriel.kasmi@minesparis.psl.eu}

\begin{document}
\maketitle

\section*{HIGHLIGHTS}
\begin{itemize}
\item A Bayesian framework to audit energy registries with remote sensing
\item A closed-form rule sets the annotation effort for a target precision
\item In France, national agreement within 3.3\% hides local gaps up to 61\%
\item Distribution-chain fragmentation explains 45\% of the core missing capacity
\end{itemize}

\section*{SUMMARY}
Tracking the energy transition requires reliable statistics on renewable deployment. Rooftop photovoltaics (PV) are especially hard to track, owing to their decentralised nature, and the resulting inaccuracies in official statistics are known but not quantified. Remote sensing offers an independent way to identify rooftop PV systems. We introduce a Bayesian framework to estimate the ground-truth rooftop PV capacity from remote sensing detections, turning an imperfect detector into an uncertainty-aware measurement instrument. Applied to France, the corrected detections estimate a capacity of 4.03 GWp [3.96--4.11] (99\% credible interval) of rooftop PV below 36 kWp, matching the transmission system operator's connection data within 3.3\% nationally, while identifying local under-reports of up to 61\% of local capacity. We also document and quantify a significant truncation bias in French rooftop PV open data. Beyond France, the approach paves the way for more reliable estimates of rooftop PV capacity worldwide.

\section*{KEYWORDS}
photovoltaics; remote sensing; energy registries; deep learning; data quality; Bayesian calibration; energy statistics

\section*{CONTEXT \& SCALE}

Rooftop photovoltaics are the most numerous type of power-generating asset, yet also the hardest to track: unlike a handful of power plants or wind farms, they are installed piecemeal by individual owners across millions of small rooftops. As a result, official statistics on how much capacity exists, and where, are uneven across countries, and their completeness is generally assumed rather than tested.

We show that satellite and aerial imagery can supply an independent measurement of rooftop PV capacity. Adapting a Bayesian framework from ecological population surveys, we turn a noisy image-based detector into a calibrated estimate with quantified uncertainty. We then use it to audit an existing registry rather than the reverse. Applied to France, the corrected estimate agrees with the national grid connection registry to within 3.3\%, but conceals sharp local gaps: registry coverage falls as low as 39\% of the estimated fleet in some regions. We trace much of this shortfall to administrative fragmentation --- reporting that crosses from a national grid operator to smaller local distributors loses data along the way ---  a mechanism that alone explains 45\% of the core missing capacity.

The same approach is applicable beyond France: wherever high-resolution imagery is available, it can check an existing registry's accuracy. Where no reliable registry exists, it can deliver the first quantified measurement of a rooftop solar fleet that has never been counted. This approach gives grid operators, statisticians, and policymakers a tool to plan for a segment of the power system that has so far escaped systematic accounting.
\section*{INTRODUCTION}
Photovoltaic installed capacity has experienced unprecedented global growth, and a significant share of it is distributed. Hundreds of thousands of small rooftop systems whose aggregate behaviour shapes grid operations through voltage excursions, reverse flows and forecasting uncertainty\cite{ipcc_summary_2021,iea_renewable_2023,laevens_observational_2021,pierro_impact_2022}. Installed capacity has been shown to be the dominant source of uncertainty in rooftop PV production estimation, ahead of meteorological and technical factors\cite{kasmi_balancing_2025,huxley_uncertainties_2022,saint-drenan_empirical_2015}. Knowing how much capacity exists, and where, is therefore an operational requirement for grid operators.

Transmission system operators (TSOs) and governmental bodies meet this requirement with administrative registries. The completeness and quality of these registries vary greatly from one country to another. Centralised systems such as France's grid-connection data and its public counterpart, the {\it Registre national d'installations} (RNI), theoretically provide comprehensive coverage\cite{kasmi_enhancing_2024}. In Germany, the Marktstammdatenregister 
requires PV owners to register directly with the regulator\cite{bundesnetzagentur_marktstammdatenregister_2022}. In the United States, the reference dataset on distributed PV is not an administrative record at all but a voluntary 
compilation from 72 organisations across 31 states\cite{barbose_tracking_2024}. Finally, in fast-growing markets such as Pakistan, official reports and independent estimates can differ several-fold. Existing assessments rely on crowdsourcing or internal consistency checks\cite{stowell_harmonised_2020,huxley_uncertainties_2022,tepe_improving_2023}, which can detect implausible records but are structurally blind to what is absent: an installation that was never registered fails no consistency test.

Remote sensing provides an independent and scalable way of mapping rooftop PV assets. Advances in deep learning\cite{lecun_deep_2015} and the growing availability of high-resolution orthoimagery have enabled detection pipelines tailored to mapping rooftop PV at the scale of entire countries\cite{yu_deepsolar_2018,malof_mapping_2019,kausika_geoai_2021,mayer_3d-pv-locator_2022,kasmi_enhancing_2024,bouaziz_high-resolution_2024,lindahl2023mapping,thebault2025comprehensive}. These methodologies produce comprehensive, harmonised datasets of rooftop PV systems using little human effort compared to crowd-sourced approaches to mapping distributed assets\cite{stowell_harmonised_2020,kasmi2026openpvmapper}. However, deep learning-based mapping remains prone to detection errors\cite{li_understanding_2021,kasmi_space-scale_2025}. It is therefore bounded to provide, at best, a noisy estimate of installed capacity. In current practice, such mapping is evaluated against official statistics\cite{rausch_enriched_2020,kasmi_towards_2022,thebault_comprehensive_2025}, implicitly treating administrative data as ground truth. The validation arrow has never been reversed at the national scale.

Viewing remote sensing-based mapping of rooftop PV installations as a noisy estimation of an underlying ground truth capacity, we introduce a Bayesian framework to retrieve such capacity. Our approach is grounded in well-established statistical methodologies: dual-system and capture--recapture estimation, which formalise how two imperfect enumerations of a population audit one another\cite{hook_regal_1992,hook_regal_1995,chao_2001}; detection-probability modelling from ecology\cite{mackenzie_estimating_2002}; and accuracy-adjusted area estimation from land-cover science\cite{card_1982,olofsson_making_2013,olofsson_good_2014}. We transpose this apparatus into a three-component audit protocol for energy registries, and evaluate our methodology in France, using the TSO grid connection data as the reference dataset to evaluate. 

The main result is a two-scale picture: nationally, the corrected estimate (4.03 GWp [3.96--4.11] (99\% credible interval) below 36~kWp) and the TSO registry agree to within 3.3\%. This national convergence shows that our method enables us to effectively retrieve a plausible ground truth PV capacity. On the other hand, at the local scale, the agreement dissolves. We find strong evidence of under-reporting in several geographical units, with registry coverage as low as 39\% of the estimated fleet. We show that 45\% of the core missing capacity in the grid connection data can be explained by the administrative fragmentation at the distribution level. We further document a truncation bias in the public data. Privacy rules remove municipalities with fewer than 10 registered systems from the public registry. This truncation leads to hiding more than 40\% of the capacity in more than 10\% of all the French departements, mostly concentrated in rural areas. This truncation does not remove  capacity from the departmental total, but it removes something else the public data cannot  recover: the fine-grained detail needed to study rooftop PV's territorial dynamics. Rooftop PV development exhibits spatial clustering that geospatial, computer-vision, and econometric analyses have repeatedly documented\cite{bollinger_peer_2012,davidson_modeling_2014,dharshing_household_2017,wang_deepsolar_2022,mehiyddine_understanding_2025}. A public dataset that erases its finest-grained observations is unable to support the analyses the territorial character of rooftop PV calls for.

Our framework extends naturally to countries where official registries are known to struggle to track rooftop PV deployment. That the corrected estimate converges with a mature registry at national scale licenses the extension: where no comparable reference exists, corrected detections can serve as a primary estimate --- with quantified uncertainty --- of the rooftop PV installed capacity.

\section*{RESULTS}

\subsection*{Estimating ground truth rooftop PV capacity to audit registries}

Rooftop PV detected on aerial imagery cannot be compared with a registry as it stands. A detection product contains false positives and misses installations (false negatives). Detections are measured as rooftop surface area and converted to an installed-capacity estimate via a fixed coefficient (Methods). The audit's first step is therefore to turn these raw detections into a corrected estimate of installed capacity, which is then used as an independent reference against which the registry can be checked.

Ecological surveys have faced a similar problem: estimating a true, unknown value from noisy observations\cite{mackenzie_estimating_2002, pollock_large_2002, royle_hierarchical_2008}. We apply the same Bayesian logic to reconstruct installed capacity from noisy detections. In a Bayesian framework, what we know about an unknown quantity before seeing any data — the prior — is combined with what the data show — the likelihood — into a posterior distribution that reflects both.

The prior is the detector's error, measured department by department. This means estimating two error rates: how many detections are false positives, and how many real installations the detector misses. We manually check a subset of the detections to estimate the former, which gives precision; separately, we check whether installations identified independently of the model appear among the detections, which gives recall. The data are the raw detections themselves. Combining the two re-weights each department's detected capacity by its measured precision and recall. This step is valid only if the error rates, measured by counting systems, are constant across installations' sizes. While this assumption is likely not to hold across the whole range of PV systems, within the $\leq$36 kWp it largely holds: recall shows no meaningful size-dependence, and a moderate precision-side deviation is carried through the audit explicitly rather than assumed away (Methods, Practical implementation). The result is a posterior distribution over the department's true installed capacity, from which we report a mean value and a 99\% credible interval.

We adopt a conservative decision rule to keep the audit's claims trustworthy: a department is flagged as under- or over-reported only when its reported capacity falls outside this 99\% credible interval. The margin is wide by design: the priority is to avoid raising false claims about the accuracy of the grid connection data, even at the cost of missing some genuine gaps.

Two conventions fix the scope of the audit. Throughout, rooftop PV means installations at or below 36~kWp --- a subset of rooftop PV corresponding to residential installations and small commercial, industrial, and agricultural rooftops. The focus on this segment is motivated by two facts: it accounts for 99\% of grid connections\cite{kasmi_remote_2024}, and it has historically carried the largest uncertainty in PV power production estimation\cite{kasmi_enhancing_2024}. Grid-connection data do not label whether a given installation is roof-mounted, ground-mounted, pole-mounted, or tracker-equipped; instead, every installation is treated as an injection point characterised by an installed capacity. The $\leq$36~kWp threshold stems from a French administrative convention that separates residential/individual installations, subject to anonymisation rules, from larger rooftop installations; a comparable logic applies elsewhere --- in Germany, for instance, the location of solar installations owned by natural persons is not published in the Marktstammdatenregister below 30~kWp\cite{bundesnetzagentur_marktstammdatenregister_2022}. The reporting unit is the \emph{department} (NUTS-3); Paris and its three inner-ring departments are pooled into a single reporting unit. For brevity, we refer to all 93 resulting reporting units, including this Paris–inner-ring aggregate, as "departments" throughout.

The Methods give the full protocol and the assumptions on which it rests. One piece remains to be described here: aligning the two calendars. Imagery and registry describe the same fleet, but at different moments in time — a system connected after the photograph was taken is not a registry gap. In France, both dates are known to the day, for every municipality and every connection, which lets us stop the registry exactly at the image acquisition date for each municipality. Note~S7 quantifies what is lost when connection dates are known only to the month.

\subsection*{The French case: national convergence, local blind spots}

Applied to continental France, the pipeline returns 589,623 raw detections totalling 2.90~GWp within the perimeter. The protocol corrects this into a national estimate of 4.03~GWp [3.96--4.11~GWp, 99\% CI] and 822,981 installations. The reference is the TSO registry, independently de-duplicated and date-matched to the day (Methods). Once aligned, it stands at 3.90~GWp so that the two estimates agree to within 3.3\%. The administrative reporting chain and our calibrated detection pipeline, sharing no input data, then converge on the same national figure. This is a full-scale sanity check, and the protocol passes it. A residual remains under the baseline specification, but it is not stable enough across the specification range to be carried as a claim (Note~S1). 

The TSO registry that this national figure rests on is itself an aggregate, built from connection data reported by distribution operators of differing institutional lineage: the incumbent DSO shares with the TSO decades of joint infrastructure inherited from the unbundling of a single, formerly vertically-integrated national utility, while other operators built and maintain their own, independent reporting channel to the TSO's information system. This heterogeneity cancels out at the national scale reported above, but does not have to at the local scale (Lessons from the French case study, below).

The local picture is different (Figure~1). 
\textbf{25 geographical units show capacity absent from the TSO's connection data at the 99\% credible level, totalling 228.2~MWp (5.7\% of the national corrected capacity). Conversely,8 geographical units display over-capacity in the grid connection data, totalling 99.4~MWp (2.5\%)}. The signal is directional as well as numerical: under-reporting geographical units carry more than twice the capacity of over-reporting ones, and the missing capacity dominates the local picture.

This baseline rests on the one assumption discussed in the previous paragraph:  detection performance does not
depend on installation size. The specification battery (Methods) tests this directly, replacing the count-based correction with a capacity-weighted one that requires no such assumption. The signal survives independently of it: even under the most conservative pairing of specifications, a
hard core of 18 geographical units and 172~MWp still remains flagged as missing PV capacity--a lower bound on the missing capacity
identified by the audit. The gaps are not marginal. In the worst-affected unit, Corse-du-Sud, registry coverage falls to 39\% (or, symmetrically, misses 61\%) of the estimated fleet.

Applying the same audit methodology to the public registry (RNI) highlights the truncation bias resulting from the privacy rule that censors municipalities with fewer than ten installations. Audited at face value, the RNI looks far less complete than the TSO registry. 32 geographical units are flagged, due to the fact that the privacy rule dropped 14,019 municipalities with detected PV from the fine-grained view. Correcting for this censoring
collapses the count to 8 geographical units, all falling within the hard core subset of the TSO registry. 

\noindent\includegraphics[width=0.48\textwidth]{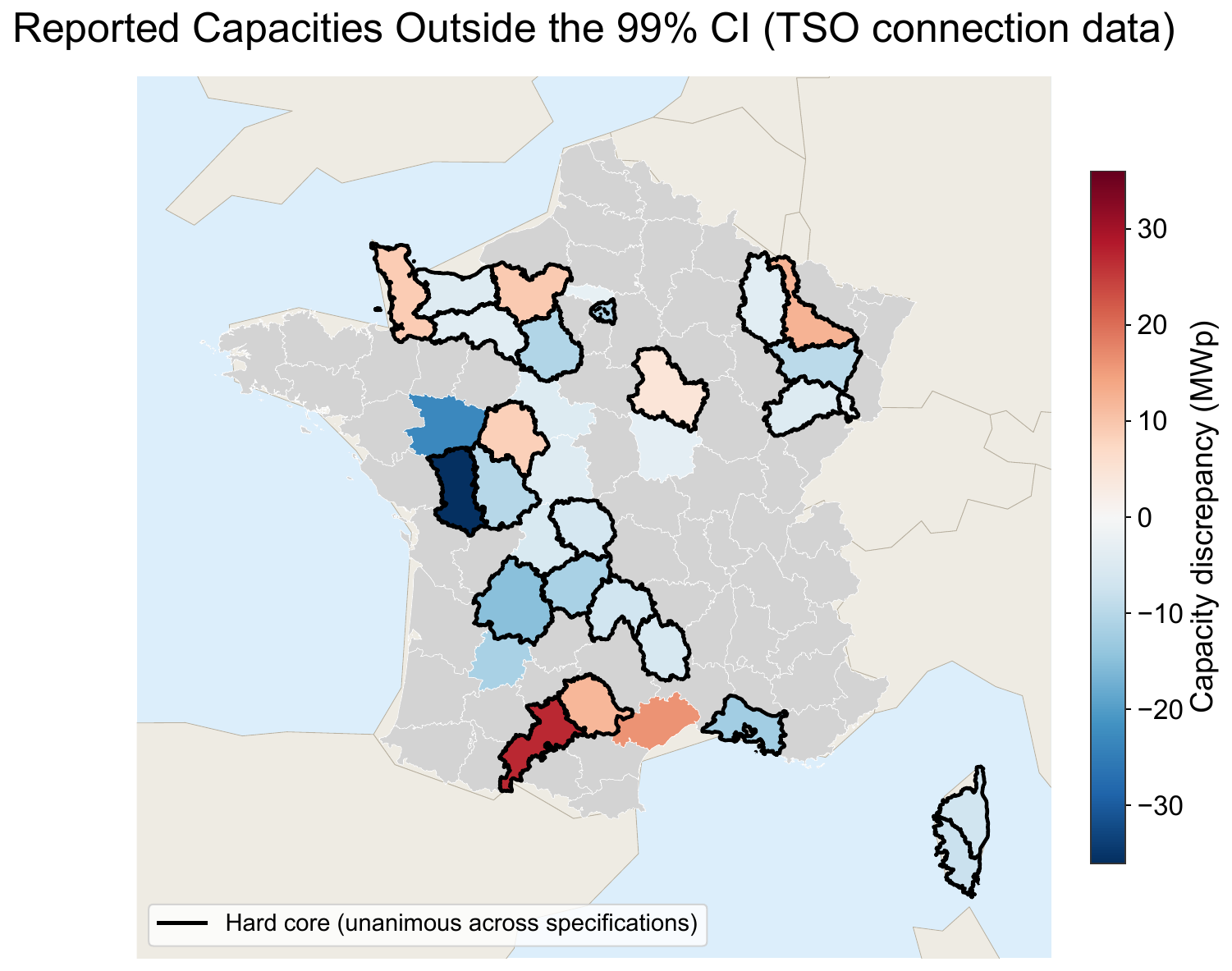}\hfill
\noindent\includegraphics[width=0.46\textwidth]{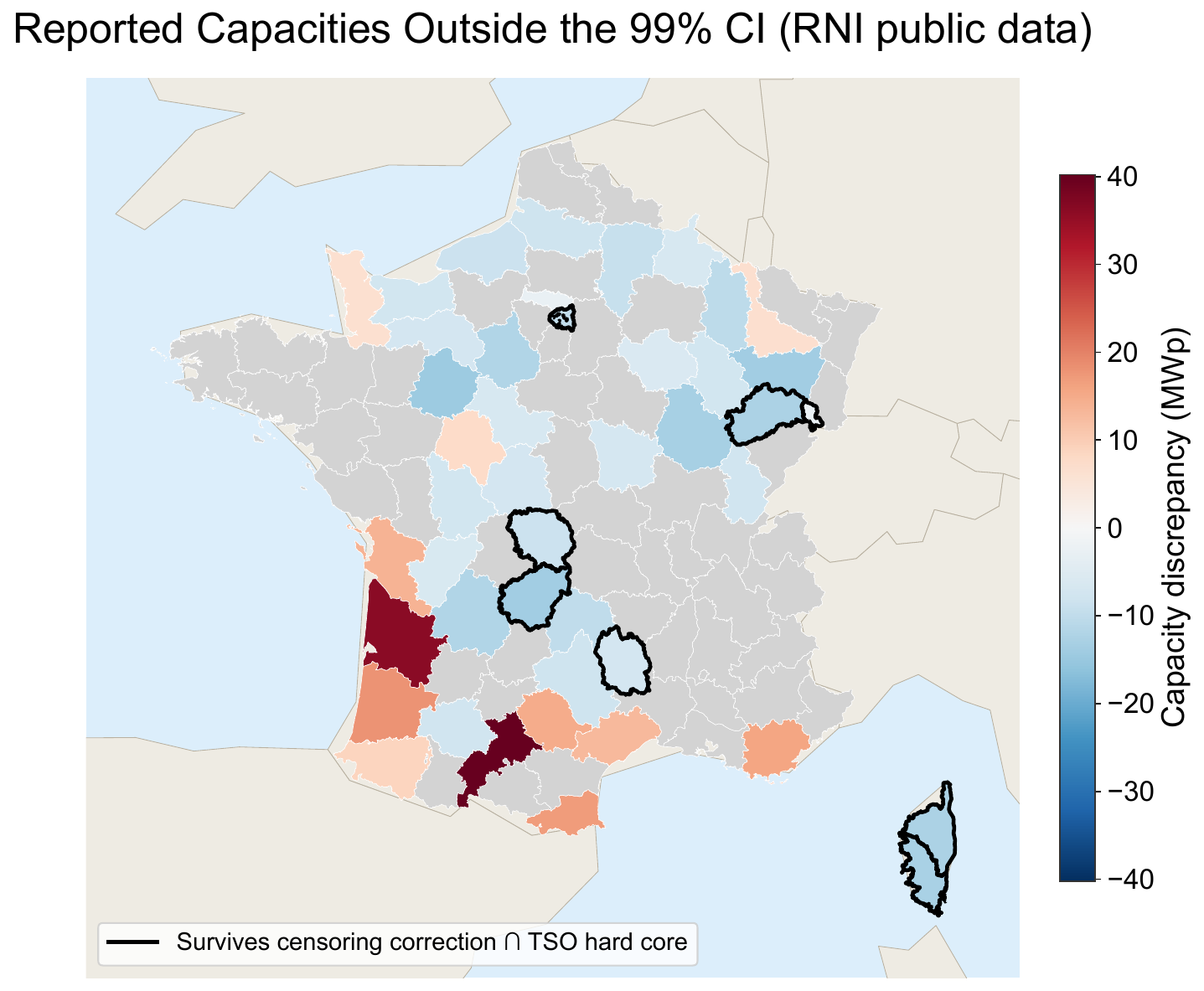}\\[4pt]
\noindent\includegraphics[width=0.95\textwidth]{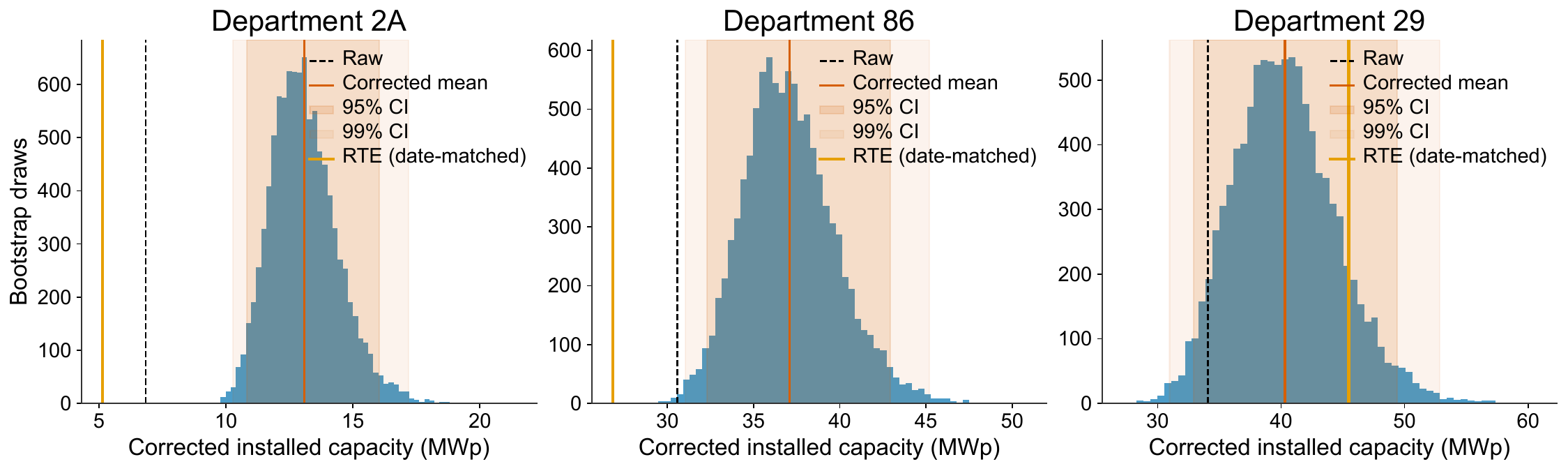}

\subsection*{Figure 1. The French case: national convergence, local blind spots.}
(A, top left) Under-reported departements against the TSO connection data (day-level alignment): geographical units whose registry value falls outside the 99\% credible interval of the corrected estimate. Bold outline: the hard core, the subset of under-reported departements that remain flagged under the specification battery (Methods), corresponding to a floor on the missing capacity identified by the audit. (B, top right) The same audit against the public registry (RNI, year-level alignment), where departmental values are reconstructed from municipal
data---exposing the truncation bias. Bold outline: the 8 departements that survive correction for censoring and fall within the TSO hard core in (A). (C, bottom) Rooftop PV capacity posteriors for two incomplete departements (Corse-du-Sud (2A), Vienne (86)) and one well-calibrated departement (Finist\`ere (29)), TSO registry value overlaid.

\subsection*{Lessons from the French case study}

Despite a centralised power system, France counts about 130 distribution system operators, one serving serving over 90\% of municipalities and about 95\% of end users. The remaining municipalities are served by local distribution companies ({\it Entreprises locales de distribution} ELDs), which report to the registry through their own channels. Whether crossing that reporting interface degrades registry completeness can therefore be measured directly.

Under the assumption that the detector's performance does not vary with the distributor (this assumption is tested in the Methods) we compare the ratio of registry to detected capacity between ELD-served and incumbent-served municipalities. We test this on the 14,551 municipalities where DeepPVMapper detects at least 50~kWp of rooftop PV, below which the coverage ratio is dominated by detection noise on a handful of roofs (results are unchanged at thresholds of 20 and 100~kWp; Robustness checks, Empirical strategy of the fragmentation mechanism). 

Municipalities served by a local operator show \textbf{42\% lower registry coverage than municipalities of the same department served by the incumbent DSO} (38\% once municipality size is controlled for). The effect is directionally consistent across every further test. Detected capacity itself does not differ between the two groups within a department, so the gap cannot be a detection artefact. Local operator municipalities are also more likely to host detected PV with no registry record at all, 2.2\% against 0.24\%, although that margin rests on 18 municipalities and is not statistically resolved. Coverage is lower in 16 of the 19 departments where both operator types are sufficiently represented (Methods; Figure~2B).

Part of this gap plausibly reflects regulatory obligations: ELDs serving fewer than 100,000 clients report connection data to RTE only twice a year, against a monthly cadence for larger distributors\cite{republique_francaise_collecte_2023}. This is not sufficient on its own, however: some of the largest ELDs, already subject to the monthly cadence given their client base, show coverage no higher than smaller ones — pointing to a broader difference in how thoroughly each operator's reporting channel is integrated with the TSO's information system, beyond reporting frequency alone (Note~S7 shows, moreover, that the gap is not a dating artefact of this reporting cadence either).

This finding lets us decompose the audit's most conservative signal of missing capacity: the 18-departements, 172~MWp hard core identified in the previous section, rather than the full 25-departements baseline, since attributing a cause to a discrepancy is only meaningful for the part of it that survives every specification. Tagged unit by unit, fragmentation accounts for \textbf{45\% of the hard-core missing capacity}, or 77 of 172~MWp---a share of the capacity itself, not of explained variance. The remaining 95~MWp, spread over 12 departements, represents 27 to 40\% of their estimated fleets (Figure~2A). Nothing we could test accounts for it: not fragmentation, not connection lag in either direction, not any other observable mechanism. Non-declaration and residual registry defects remain open possibilities---hypotheses for the operator to pursue on the registry side, since the audit has exhausted what can be established from the measurement side.

\noindent\includegraphics[width=0.95\textwidth]{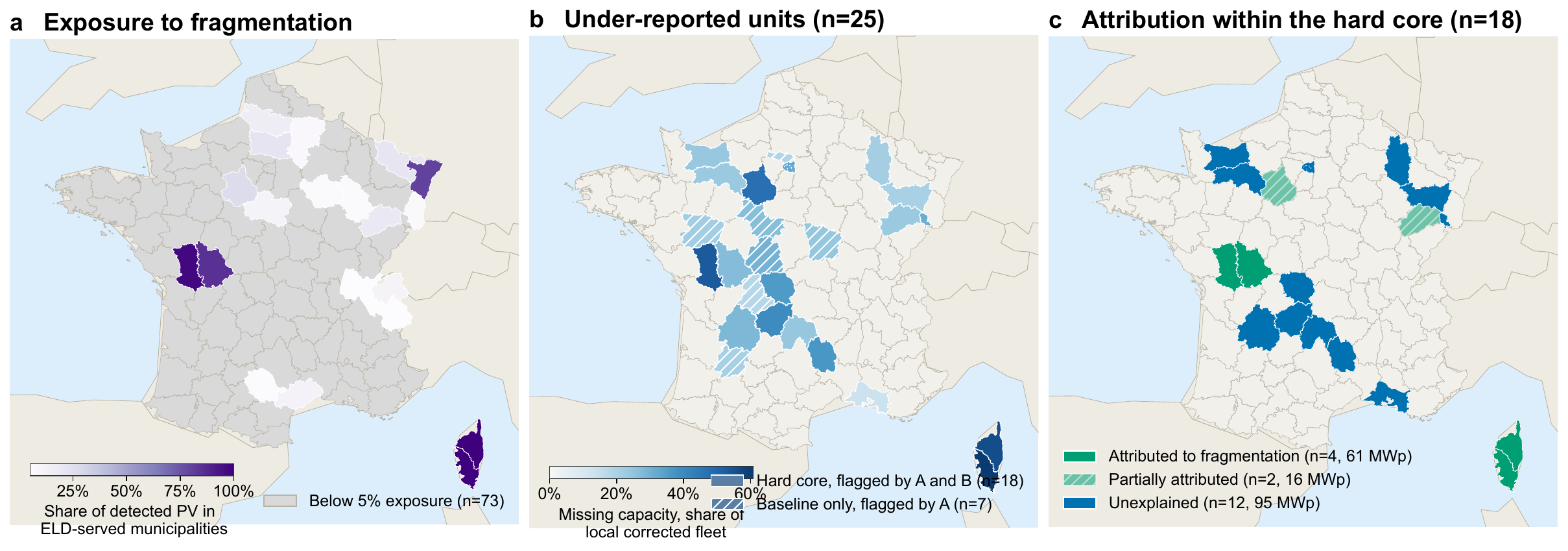}\\[4pt]
\noindent\includegraphics[width=0.85\textwidth]{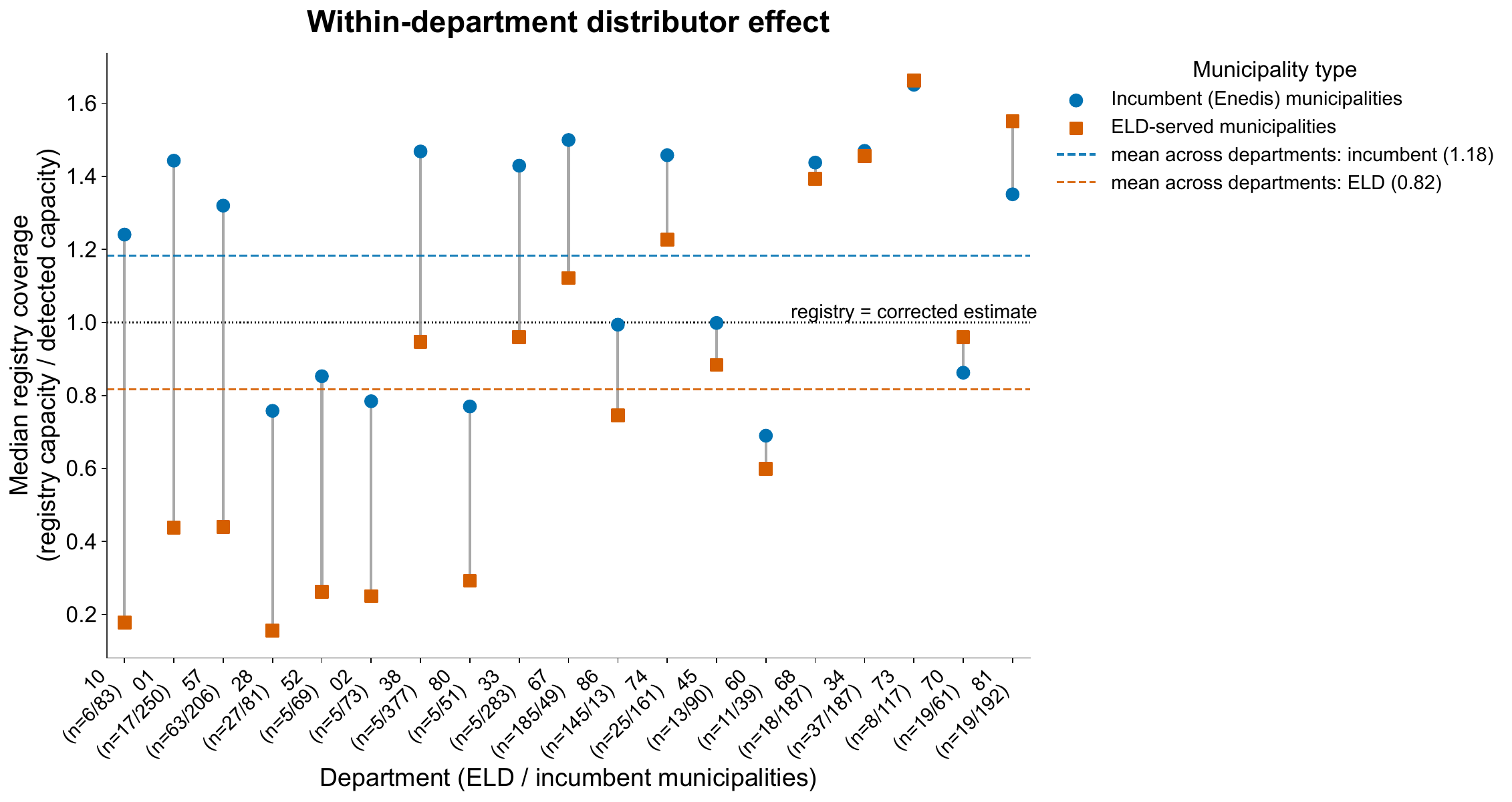}
\subsection*{Figure 2. The impact of administrative fragmentation on registry completeness.} 
(A, top left) Exposure to fragmentation: share of detected PV capacity located in municipalities served by a local distribution company (ELD) rather than the incumbent, by department ($n=20$ departments above a 5\% exposure floor; the remaining 73, shown in grey, have negligible ELD exposure). (B, top centre) Under-reported departments at the baseline specification ($n=25$): missing capacity as a share of the department's corrected fleet, distinguishing the hard core --- departments flagged under both specifications \textbf{A} and \textbf{B} ($n=18$) --- from those flagged under the baseline specification \textbf{A} alone (hatched, $n=7$). (C, top right) Attribution within the hard core ($n=18$): departments where fragmentation fully accounts for the missing capacity (green, $n=4$, 61~MWp), partially accounts for it (hatched green, $n=2$, 16~MWp), or where the gap remains unexplained (blue, $n=12$, 95~MWp). (D, bottom) The fragmentation effect: median registry coverage (registry capacity / detected capacity) in ELD-served versus incumbent-served municipalities of the same department, restricted to municipalities with at least 50~kWp of detected capacity ($n$ per department given as ELD/incumbent counts); coverage is lower in ELD-served municipalities in 16 of 19 departments (Wilcoxon $p = 0.0003$).

\section*{DISCUSSION}

\subsection*{Summary of the findings}

How can we estimate a ground truth rooftop PV capacity from imperfect measurements? This work shows that a Bayesian detection-probability framework, long used to reconstruct true abundance from imperfect ecological counts, can be adapted to answer this question. Extending the existing logic to this new field entails taking into account three new requirements for the estimation to be valid. First, the observer is not a person but a multi-stage deep learning pipeline, and its usable error rate is not its benchmark accuracy but what survives every downstream processing step --- on our data, recall alone falls from 0.84 to 0.61 between the two. Second, the object under audit is not a natural population but the record of an institution, so the corrected estimate has to be defensible by design, not merely accurate: every correction we apply is direction-agnostic, its sign set by the data rather than by us. Third, the quantity of interest is not a discrete count but a continuous installed capacity, which requires an explicit test of whether detection performance holds evenly across installation sizes before any correction can be trusted. Answering these three questions turns a noisy PV detector into a calibrated, uncertainty-aware measurement instrument, rather than the validation target it has always been treated as.

Applied to France, this instrument provides, to our knowledge, the first full-scale evaluation of a national grid connection registry's completeness. The result is a national convergence that conceals a local picture predominantly oriented toward under-reporting: two independent instruments agree at the scale where system operation aggregates, and diverge sharply below it.

Against the TSO registry, these local gaps are substantively real, and a difference in how thoroughly each operator's reporting is integrated with the TSO's information system --- rather than reporting frequency or operator diligence --- is their principal identified driver. This is, to our knowledge, the first empirical evidence that crossing operator reporting interfaces degrades registry completeness. The public registry inherits these same gaps, since it is built from the TSO's own data, but a second, unrelated mechanism sits on top of them: a publication rule censors capacity below a reporting threshold, hiding it from fine-grained view without removing it from the fleet. This affects 14,019 municipalities, and it is purely an artefact of how the data is published, not a further sign of missing capacity.

For data users, the consequences are structural. Planning at the local scale against the public registry structurally misses a sizeable share of PV capacity, concentrated in rural areas---the same areas where hosting-capacity constraints and voltage management are most sensitive to knowing the true fleet. Benchmarking detection models against the public registry absorbs both censoring and dating artefacts into reported model error, understating how well such models actually perform. Beyond the registry itself, correcting installed capacity has downstream value: capacity is the dominant source of uncertainty in rooftop PV production estimation, ahead of meteorological and technical factors\cite{huxley_uncertainties_2022,kasmi_balancing_2025}, and this uncertainty compounds where local under-reporting is largest.

\subsection*{Limitations of the study}

Our approach has several limitations. Where possible, we addressed them through wider uncertainty margins; the rest are left for future work to resolve explicitly. First, the audit is built on binary PV detection: solar-thermal panels are visually close to PV, and a confusion shared by the detector and annotators would be invisible to our label-reliability checks; the pipeline's detection floor ($\approx$11 m$^2$) excludes the dominant small thermal formats and, symmetrically, closes the question of small kilowatt kits (or plug-and-play installations). Nevertheless, a marginal contribution of thermal units cannot be fully excluded. Such a contamination would inflate, not deflate, the corrected installed capacity, since it raises both the raw detected capacity and the measured precision used to correct it. This risk is tempered by the national convergence itself: a systematic inflation of this kind would be expected to widen, not narrow, the 3.3\% gap between the corrected estimate and the TSO registry. 

Second, the recall ground truth is seeded by OpenStreetMap, whose coverage is not spatially random\cite{kasmi2026openpvmapper}: it passes a source-concordance test (at constant department, recall on OSM points is indistinguishable from recall on manual annotations; Note S2), but a residual intra-rural selection cannot be tested and is acknowledged. 

Third, the surface-to-capacity conversion is the audit's largest single sensitivity: it is bracketed (5.0--6.0~m$^2$/kWp), anchored on the registry's own implied coefficient in well-calibrated departements (5.63~m$^2$/kWp vs 5.5~m$^2$/kWp retained), and every departement whose status depends on it is flagged rather than claimed. Fourth, remote-sensing detection identifies the physical presence of PV-like material on a rooftop, not its electrical connection status. Beyond the small plug-and-play kits already excluded by the detection floor (above), a detected, correctly-sized installation could still be one exempt from standard connection declaration, an off-grid or autonomous system never intended to connect, or a decommissioned unit left physically in place. We cannot fully exclude a marginal contribution of such cases to the corrected detected capacity, which means the local gaps identified here should not be read as certifying that all of the corresponding capacity is genuinely missing from the registry through under-reporting. Fifth, the protocol's temporal component was demonstrated under gold-standard dating; its robustness to degraded date granularity is quantified in Note~S7 and conditions the export of the audit to registries with coarser dates. Sixth, both the registries audited here and the protocols that feed them are themselves evolving: RTE, distribution operators, and the regulator are engaged in ongoing discussions to revise reporting-frequency and aggregation rules. The results reported here describe the state of these registries over the audit's imagery and data vintage (2022–2025); they should be read as a measurement at that point in time, not as a definitive or permanent characterisation of any operator's practice.

\subsection*{Broader applicability}

Beyond the French case, the methodology is particularly relevant to poorly monitored, fast-growing emerging markets. Its ingredients are generic: orthoimagery, and a bounded annotation effort whose cost scales with the number of reporting geographical units rather than with territory. The budgeting rule further discussed in the Methods lets any prospective audit price its precision in advance, before a single label is drawn. The estimator itself is released as a standalone package, \texttt{bayesian-pv-census}\cite{kasmi_bayesian-pv-census_2026}, which is agnostic to the application domain: a reporting unit is any entity with a raw total and a validation sample. Transposing the audit to another country therefore requires new data, not a new statistical implementation.

The mechanism identified in France --- information loss where reporting crosses an operator interface --- makes a falsifiable prediction about registries elsewhere. Systems whose completeness depends on operator-mediated reporting chains, but at far higher fragmentation than France's 130 distribution operators, should exhibit the effect at larger amplitude: the United States' patchwork of 72 reporting organisations is the clearest candidate. Conversely, registries built on direct owner registration bypass the mediated chain entirely and should behave as a natural control regardless of how fragmented the underlying distribution network is; Germany's Marktstammdatenregister, which registers PV owners directly rather than through their distribution operator, is one such case despite the country's 850+ network operators\cite{bundesnetzagentur_marktstammdatenregister_2022}. The protocol introduced here is precisely the instrument required to test this prediction.

Pakistan illustrates the opposite extreme --- not a registry that misreports part of the fleet,  but a segment with untrustable administrative census. It shows the scale  such gaps can reach: the national net-metering registry lists on the order of 5--6 GW of connected capacity\cite{nepra_state_2025}, while independent estimates combining satellite mapping and household-survey data place actual distributed solar deployment at 27.5--33.5 GW\cite{transitionzero_shedding_2026}---a five-to-sevenfold gap, accumulated within roughly two years, in a segment with no scheduled administrative census to correct it. In such cases, corrected remote sensing is not an audit of the registry; it is the only measurement.

\section*{METHODS}
\subsection*{Detailed audit methodology}

Our aim is to provide an unbiased and with bounded uncertainty estimate of the rooftop PV capacity in a given geographical unit. To this end, we leverage the detections of a rooftop PV detector, deployed on aerial orthoimagery. Our underlying logic for correcting the detections is Bayesian. Bayes' theorem provides a framework for updating prior beliefs about a parameter $\theta$ in light of observed data $(x_1,\ldots,x_N)$. The posterior distribution of the parameter is given by
\begin{equation}
    p\left(\theta \mid x_1,\ldots,x_N\right) = \frac{
p\left(x_1,\ldots,x_N \mid \theta\right) p(\theta)
}{
p\left(x_1,\ldots,x_N\right)
},
\end{equation}
where, $p(\theta)$ denotes the {\it prior distribution}, representing beliefs about the parameter before observing the data. The term $p(x_1,\ldots,x_N\mid\theta)$ is the {\it likelihood}, quantifying how compatible the observed data are with a given value of $\theta$. The resulting $p(\theta\mid x_1,\ldots,x_N)$ is the {\it posterior distribution}, the updated belief about $\theta$ after observing the data. Since the marginal likelihood does not depend on $\theta$, this is equivalently written, up to a normalising constant, as
\begin{equation}
p\left(\theta \mid x_1,\ldots,x_N\right) \propto p\left(x_1,\ldots,x_N \mid \theta\right)p(\theta).
\end{equation}

Unlike existing work in ecology\cite{hook_regal_1992,card_1982,olofsson_good_2014}, surveyed in Note~S9, we do not aim to estimate this posterior in full generality. We wish to derive a posterior distribution of the true capacity since this distribution gives us a mean value and a variance. We show that under a set of assumptions, we can interpret this mean value as the true capacity. The variance, on the other hand, can be used to estimate the uncertainty of the instrument and make statements such as knowing the capacity up to a given percentage.

Our raw, assumed-to-be-biased estimation of the rooftop PV capacity is the output of a deep learning pipeline deployed to detect installations. To unbias the detections, we need to correct this measurement by the detection errors of the deep learning model, namely its precision and recall. The precision and recall provide the update to the prior capacity estimation and yield the posterior distribution on the capacity. Concretely, the correction is
\begin{equation}
C_{\text{adj}} \;=\; \frac{P}{R} \times C_{\text{raw}},
\end{equation}
where $C$ denotes rooftop PV installed capacity, and $P, R$ are precision and recall, estimated from the pipeline's raw detections. $C_{\text{adj}}$ is only interpretable as the true rooftop PV capacity, $C_{\text{true}}$, under four assumptions, later referred to as the {\bf estimation assumptions}.

\begin{assumption}[\textbf{H1} --- Homogeneity]
Detection status is independent of installation capacity: true positives, false positives, and false negatives have the same mean capacity.
\end{assumption}
\begin{assumption}[\textbf{H2} --- Exhaustive geographic coverage] The detector is run over the entire reporting unit; no sub-region is excluded.
\end{assumption}
\begin{assumption}[\textbf{H3} --- Representative sampling]
The samples used to estimate $P$ and $R$ are drawn representatively from, respectively, the full set of raw detections and the true population of installations.
\end{assumption}
\begin{assumption}[\textbf{H4} --- Unbiased capacity conversion]
For a correctly detected installation, its estimated capacity is unbiased for its true capacity.
\end{assumption}

\begin{theorem}
Under \textbf{H1}--\textbf{H4}, the correction $C_{\text{adj}} = (P/R)\,C_{\text{raw}}$ equals $C_{\text{true}}$ exactly at the true values of $P$ and $R$; and given \textbf{H2}--\textbf{H3}, the estimator $\hat{C}_{\text{adj}} = (\hat P / \hat R)\,C_{\text{raw}}$ is a \emph{consistent} estimator of $C_{\text{true}}$, converging as the annotation samples $n_P, n_R \to \infty$. 
\end{theorem}

The proof of Theorem 1 is given in Note~S1. 

We make three modelling choices in this work, for which our methodology is not tied : 

\begin{enumerate}
    \item We focus on the rooftop PV power capacity below 36 kWp. This is motivated by the fact that this segment is the least observable in the French connection data. 
    \item We focus on installed capacity rather than the number of installations. Capacity in kWp (rather than surface area) is the common unit in which energy system practitioners characterize energy statistics. Additionally, in the French setting, installation counts are subject to substantial definitional uncertainty. Aggregating deployment in terms of capacity avoids counting ambiguities and is consistent with standard practice among practitioners. 
    \item We stratify the proposed framework at the level of the French {\it départements}, which correspond to NUTS-3 regions in the European statistical classification. Accordingly, each capacity estimate $C_{\text{true}}$ should be understood as $C_{\text{true}}^d$, where $d$ denotes a French {\it département}.
\end{enumerate}

The framework is not intrinsically tied to these choices. The capacity threshold can, for example, be extended to include larger installations, although this would likely violate {\bf H1}. Conversely, if installation counts are used as the quantity of interest rather than capacity, {\bf H4} is no longer required. Likewise, the choice of spatial granularity is not inherent to the methodology. We use the {\it département} as the target spatial unit because it provides a suitable balance between annotation availability and the ability to capture local patterns in rooftop PV deployment. The same corrections can be applied at coarser scales, such as the regional or national level, or at finer spatial scales, provided that a sufficient number of annotations is available for each unit (Component~1, below).

Figure S1 summarizes the audit methodology, from the initial surveying using a deep learning-based detector to the definition of a decision rule to flag discrepancies in geographical units. The remainder of the section addresses this construction in two parts. The first, immediately below, shows how to estimate $P$ and $R$ so that \textbf{H3} holds, and how to use these estimates to derive $C_{\text{true}}$ from $C_{\text{raw}}$: this establishes that the corrected estimate is right on average. Being right on average is not, by itself, useful: an unbiased estimate that is also extremely imprecise cannot distinguish a real registry gap from noise. The second part, \emph{Accuracy and uncertainty}, addresses this remaining question --- how tightly the corrected estimate concentrates around its true value, and how that precision can be budgeted for in advance.

\noindent\includegraphics[width=0.95\textwidth]{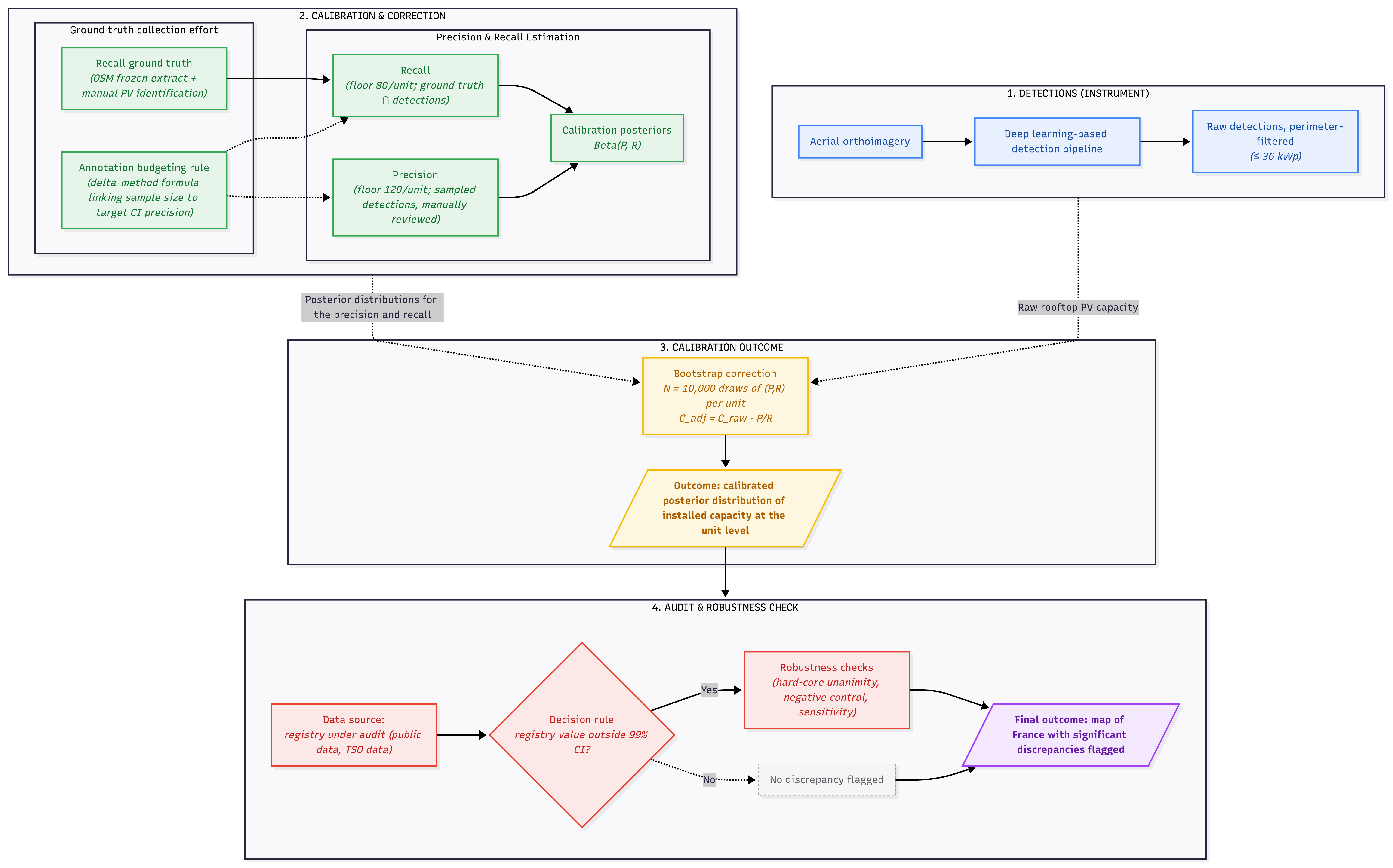}
\subsection*{Figure S1. Flowchart of the audit protocol.}
The audit proceeds in four stages.
\textbf{(1) Detections (instrument):} aerial orthoimagery is processed by a deep-learning detection pipeline; raw detections are perimeter-filtered to the $\leq$36~kWp segment analyzed in this study. \textbf{(2) Calibration \& correction:} ground truth for recall is collected independently from the model's detection. Precision is evaluated by sampling the pipeline's detections. Sample sizes can be estimated in advance via a delta-method annotation budgeting rule. These annotations are fed into Beta distributions to estimate a distribution over the precision and recall values. \textbf{(3) Calibration outcome:} the raw detected capacity (from stage 1) and the $P$,$R$ posteriors (from stage 2) are propagated through $N$ bootstrap draws of the correction $C_{\mathrm{adj}} = C_{\mathrm{raw}} \cdot P/R$, yielding a calibrated posterior distribution of installed capacity at the unit level. \textbf{(4) Audit \& robustness check:} this distribution is compared against the registry under audit; a unit is flagged if the registry value falls outside the 99\% credible interval and is carried through robustness checks (hard-core unanimity, negative control, sensitivity) before contributing to the final map of significant discrepancies, while units within the interval are retained as consistent with the registry. Throughout, the corrected estimate is interpretable as the true installed capacity only insofar as assumptions H1--H4 hold (\emph{Detailed audit methodology}); these are established against the annotated data at earlier points in the procedure, not as a discrete step in this pipeline.

\subsubsection*{Estimation of the true rooftop PV capacity} 

\paragraph{Component 1: estimation of the precision and recall}

Turning counts into a capacity estimate needs the confusion matrix of the \emph{deployed pipeline}, not the metrics reported at training or test time (Note~S8 discusses that distinction). We use the same vocabulary --- precision, recall, F1 --- for both, but they are not the same quantities: pipeline precision and recall are measured on the final, post-processed detections, after every filtering step the raw model output goes through. On our data, model-level recall reaches 0.84; pipeline recall is 0.61. A correction built on the model's own metrics would understate the true error by some 40\%, which is why precision and recall are measured directly, manually, on the pipeline's actual output.

To retrieve initial samples, we proceed separately for each quantity. Within each departement, we randomly sample $N_p$ predictions and manually review them to estimate the precision. For the recall, we construct a set of $N_r$ PV systems and see whether they can be found in the model's detections. As an additional requirement for the recall, we make sure that the detections could have been seen by the model, i.e., we discard PV systems that have been added after the image acquisition date. 

For each departement, the review of the detections yields counts for the true positives $TP$ and the false positives $FP$ and the intersection of the recall points with the detections yields a count of true positives and false negatives $FN$. In both cases, it corresponds to a Binomial count, and each combines with a Beta prior into a closed-form posterior:
\begin{equation}
P \sim \mathrm{Beta}(\alpha_P, \beta_P) \;\longrightarrow\;
P \mid \text{data} \sim \mathrm{Beta}(\alpha_P + \mathrm{TP}, \, \beta_P + \mathrm{FP}),
\end{equation}
\begin{equation}
R \sim \mathrm{Beta}(\alpha_R, \beta_R) \;\longrightarrow\;
R \mid \text{data} \sim \mathrm{Beta}(\alpha_R + \mathrm{TP}, \, \beta_R + \mathrm{FN}).
\end{equation}
The prior can be uninformative --- Jeffreys, $\alpha=\beta=0.5$ --- or it can pool information across departments. We discuss the choice of the prior in the practical implementation. These distributions correspond to distribution from which we can sample values for the precision and the recall, which are then propagated further into the pipeline to estimate the capacity. Note that this framework applied a Bayesian correction to the precision and recall themselves. 

\paragraph{Component 2: from precision and recall to a capacity posterior.}

Once $P$ and $R$ have a posterior, we propagate it by simulation. At each of $N$ draws, we sample $P$ and $R$ from their posteriors and apply the resulting ratio to the raw detected capacity:
\begin{equation}
C_{\text{adj}}^{(i)} \;=\; C_{\text{raw}} \cdot \frac{P^{(i)}}{R^{(i)}}, \qquad i = 1, \dots, N.    
\end{equation}

Across the $N$ draws, this gives a full distribution over the department's corrected capacity, from which we report a mean and a credible interval (the concrete draw count and interval levels are given below, in Practical implementation). The same factor is applied to the raw installation count. Four assumptions are needed for the corrected estimate to converge to the true capacity value.

First, detection performance should not depend on installation size (\textbf{H1}). Verifying this is not trivial, given the strong heterogeneity of PV installations: rooftop PV specifically is known to be harder to detect than utility-scale, ground-mounted installations\cite{yu_deepsolar_2018}. However, we have reason to expect it holds here: the audit is restricted to installations at or below 36~kWp, a relatively homogeneous population. We do not rest on this argument alone: the assumption is tested directly on the annotated data. Recall shows no meaningful size-dependence ---capacity-weighted pipeline recall (0.649) is statistically indistinguishable from its count-based counterpart (0.645). Precision shows a moderate deviation (0.733 capacity-weighted, against 0.766 count-based), which we do not assume away: it is carried through the audit explicitly, via the capacity-weighted specification of the battery (Robustness checks, specification battery below).

Second, the geographic coverage of the raw detections should be exhaustive (\textbf{H2}), and the samples used to estimate $P$ and $R$ should themselves be geographically representative (\textbf{H3}). We manually verified \textbf{H2} by reviewing that all tiles of all departments were mapped when constructing the final detection layer. \textbf{H3} holds automatically on the precision side: since the precision sample is drawn from the detections themselves, once \textbf{H2} is satisfied, a random sample of detections is geographically representative by construction. Recall is different: its ground truth is drawn from an external source (OpenStreetMap or manual annotation, see Practical implementation), which could in principle cluster geographically within a department. We therefore visually inspected the spatial distribution of recall points across departments and found no visible clustering (Figure S3).

Third, the estimation also rests on the surface-to-capacity conversion coefficient being unbiased (\textbf{H4}). We validated an average coefficient by comparing the capacity distributions implied by our raw detections against the TSO registry's own capacities, restricting the comparison to departments where the two already agree --- avoiding the circularity of validating the instrument against the very quantity it is meant to evaluate. The resulting coefficient (5.5~m$^2$/kWp) is close to this data-driven value (5.63~m$^2$/kWp), and slightly below the value traditionally used in the literature (6.0~m$^2$/kWp)\cite{rausch_enriched_2020}, plausibly because our coefficient absorbs additional sources of uncertainty --- the gap between a detected polygon and the true module surface, and variation in panel inclination --- that a literature-wide average does not. Full detail on this non-circular anchoring is given in Robustness checks, Surface-to-capacity conversion.

\paragraph{Component 3: temporal alignment and comparison rule.}

Imagery and registries describe the same fleet at different moments. Two failure modes follow if dates are not aligned: an installation mapped only after the aerial survey is not a genuine false negative, and capacity connected only after the survey is not under-reporting. In France we align both sides to the day: OpenStreetMap edit metadata dates the ground truth, and the TSO's connection data report the installation date up to the day. On the other hand, we have the acquisition date up to the day. We match the days at the municipality level (implementation details below; Note~S7 quantifies what is lost when only coarser dates are available).

Once the corrected posterior and the date-matched registry value are both in hand, the decision rule is simple: a department is flagged if the registry value falls outside the 99\% credible interval of its corrected distribution. This interval is deliberately wide: the goal at this stage is to demonstrate that the method works and find genuine signal, not to flag every possible discrepancy. The reported gap, where one is found, is computed against the posterior mean.

\subsubsection*{Uncertainty of the rooftop PV power estimation} 

Uncertainty on the corrected capacity is expressed as an interval width, in percent, relative to the posterior mean --- e.g., an estimate of 100~GWp $\pm$10\% means the true value is credibly within 90--110~GWp. This section derives where that width comes from, and how it can be controlled. 

\paragraph{From detection error to capacity uncertainty.} The correction $C_{\text{adj}} = C_{\text{raw}} \cdot P/R$ is a ratio of two independently estimated quantities: $P$ and $R$ come from separate annotation campaigns (Estimation of the precision and recall, above), so their errors do not covary. By the delta method, the relative uncertainty of the ratio is approximately the sum of the relative uncertainties of $P$ and $R$ individually --- each estimated, in turn, from a binomial count of size $n_P$ or $n_R$. This gives the closed-form half-width of the credible interval on a department's corrected capacity,
\begin{equation}
\pm \;=\; z\sqrt{\frac{1-P}{P\,n_P} + \frac{1-R}{R\,n_R}},
\end{equation}
where $z$ is the credible-level quantile (delta method on the posterior ratio\cite{oehlert_note_1992}. One property falls directly out of this construction: the more reliable the detector --- the closer $P$ and $R$ are to 1 --- the narrower the interval, and the more power the audit has to tell a real registry discrepancy from noise. The same holds for annotation effort: larger $n_P, n_R$ shrink the interval or a detector of given reliability.

\paragraph{Budgeting the annotation effort.} Read forward, this relation predicts how precise the audit will be, for a given detector and a given annotation sample. Inverted, it becomes a budgeting rule: it returns the annotation effort a target precision requires, before any labelling begins. Acquiring and validating precision and recall samples is the most time-consuming step of the audit, so this inversion matters practically, not just formally: it lets the annotation effort be kept to the minimum required for a chosen level of accuracy, rather than annotated in excess or, worse, under-annotated without any way to know it in advance. We use the rule in both directions throughout this paper.

\noindent\includegraphics[width=0.55\textwidth]{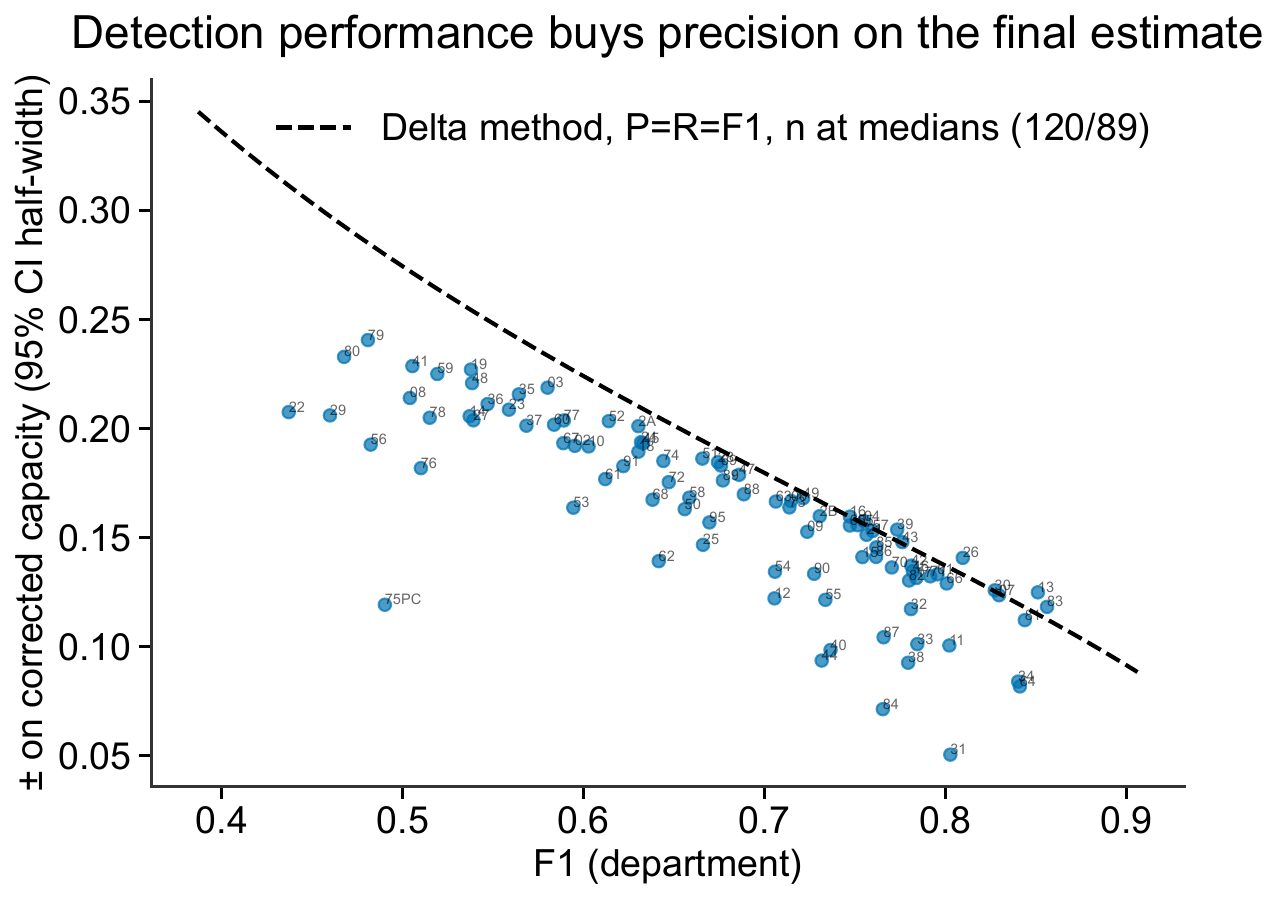}
\subsection*{Figure S2. Correlation between pipeline accuracy (F1 score) and relative uncertainty of the posterior PV capacity distribution.}
Relative half-width of the 95\% credible interval on corrected capacity, per department, against departmental F1. The dashed curve is the zero-parameter delta-method prediction, evaluated at the median annotation counts across departments (120 precision annotations, 89 recall points) and under the simplifying assumption $P = R = F1$. 
Departements sit at or below this curve because we targeted the annotation effort towards the weakest-performing locations.

\subsection*{Practical implementation for the French case study}

\subsubsection*{Collection of the precision and recall samples}

To fix the annotation budget in advance, we started from a back-of-the-envelope calculation. Using the model's test-set metrics\cite{kasmi_towards_2022} ($P = 0.81$, $R = 0.96$) and the binomial approximation $\sigma_p \simeq \sqrt{p(1-p)/n}$, initial floors of 120 precision annotations and 80 recall points per department were expected to yield roughly $\pm7$ percentage points of uncertainty on precision and $\pm4$ points on recall (95\%). Propagating these through the capacity correction via the delta method, assuming independent estimates of $P$ and $R$, gave a target relative uncertainty of approximately $\pm10\%$ on the corrected capacity --- corresponding to a labelling budget of roughly 19,200 points across France's 96 metropolitan departments. These 96 are later merged into the 93 reporting units used throughout the rest of this paper: Paris and its three inner-ring departments --- individually small and otherwise thinly annotated --- are pooled into a single unit specifically to keep the uncertainty on that unit's estimate under control, rather than leaving each of the four with its own, poorly constrained posterior.

\paragraph{Collection of the precision samples}

For precision, we sampled 120 detections per department as an initial effort (11,520 points in total), then directed additional annotation toward departments where measured precision sat closest to 0.5 --- the point of maximum uncertainty for a binomial proportion --- following the budgeting rule above. This brought the final precision campaign to 18,023 annotated detections. On average, we collected 193 points (median of 120 points) per departement to estimate the precision, with values ranging from 120 to 808 points (which corresponds to the sum of the points in Paris and its inner ring). 

\paragraph{Collection of the recall samples}

Recall ground truth combines two sources: OpenStreetMap (OSM) and a manual identification of PV systems from aerial imagery. OSM's role here is limited to providing, efficiently, a sample of installations independent of the detection algorithm against which to measure recall --- it is not treated as an independent registry competing with, or superior to, the RNI. Because the study is restricted to installations $\leq$36~kWp, which the RNI reports only in aggregated, municipality-level form rather than individually, OSM contributors could not have sourced these specific points from the RNI, unlike larger, individually-listed installations; the risk of circularity between the two sources is correspondingly low. From a frozen OSM extract of PV-related objects, we excluded all objects above the 36~kWp perimeter, applying the threshold conservatively --- discarding borderline objects rather than risking their inclusion --- and all objects certainly mapped after the relevant image acquisition date. After deduplication of points referring to the same physical rooftop, this yielded 8,457 OSM points. Departments still below the 80-point recall floor after OSM alone were supplemented with 5,563 manually added points, manually retrieved from IGN aerial imagery. Combining both sources and deduplicating a final time at the 25~m radius used throughout this protocol left 13,847 ground-truth points, with every department above its 80-point floor. Recall is then computed by intersecting these points with the raw detections layer. On average, we collected 148 points to estimate the recall, with values ranging from 80 to 1921 points and a median of 89 points. 

\paragraph{Estimation assumptions.} Overall, 31,870 points were used to estimate the precision and recall of the detections and to correct the capacity estimates, with an average of 343 samples per departement. 

Once the labels were collected, we tested \textbf{H1} directly on the annotated samples. Alongside the usual count-based precision and recall, we compute capacity-weighted counterparts: each annotated detection or ground-truth point is weighted by its estimated kWp rather than counted as one unit. Because the surface-to-capacity coefficient appears in both the numerator and denominator of these weighted ratios, it cancels exactly, making this test orthogonal to the conversion-coefficient sensitivity addressed separately in the specification battery. Weighted recall is statistically indistinguishable from its count-based counterpart (0.649 against 0.645, computed on ground-truth points carrying a surface — way objects rather than points; point-only ground truth is excluded from this specific test but remains part of count-based recall), indicating no meaningful size-dependence: smaller systems are missed more often, but contribute little capacity when they are, and the two effects cancel by construction. Weighted precision shows a moderate, one-directional deviation (0.733 against 0.766) --- false positives are, on average, somewhat larger than true positives, plausibly reflecting large reflective surfaces such as metal roofing. 

This deviation in the precision motivates the definition of a hard-core of units in our methodology to see which units are still flagged after weighting the precision and recall by the capacity (see specification B in the robustness checks). The full derivation of both weighted metrics, and the specification-battery treatment of the precision-side deviation, is given in Note~S1.

The second requirement is that they should be spatially representative of the study area (\textbf{H3}). For the precision points, this condition follows \textbf{H2} (i.e., the geographical coverage of the detector is complete). On the recall side, geographic representativeness cannot be assumed the same way, since the ground truth is drawn from an external source that could in principle cluster within a department even if detection coverage itself is complete. We therefore extensively inspected that the recall points are well spread over the departements. Figure~S3 illustrates the localisation of the precision and recall points. 

\paragraph{Annotation quality.} We manually re-annotated a sample of the validation points, blind to the original labels, to assess their reliability. Precision labels reproduce well (Cohen's $\kappa = 0.864$ on 500 relabelled points); the recall ground truth contains a small, quantified rate of invalid points (6.8\% on 250 re-reviewed points), a correction that raises, rather than lowers, measured recall. Full detail is given in Note~S2 (Ground truth validity).

\noindent\includegraphics[width=0.85\textwidth]{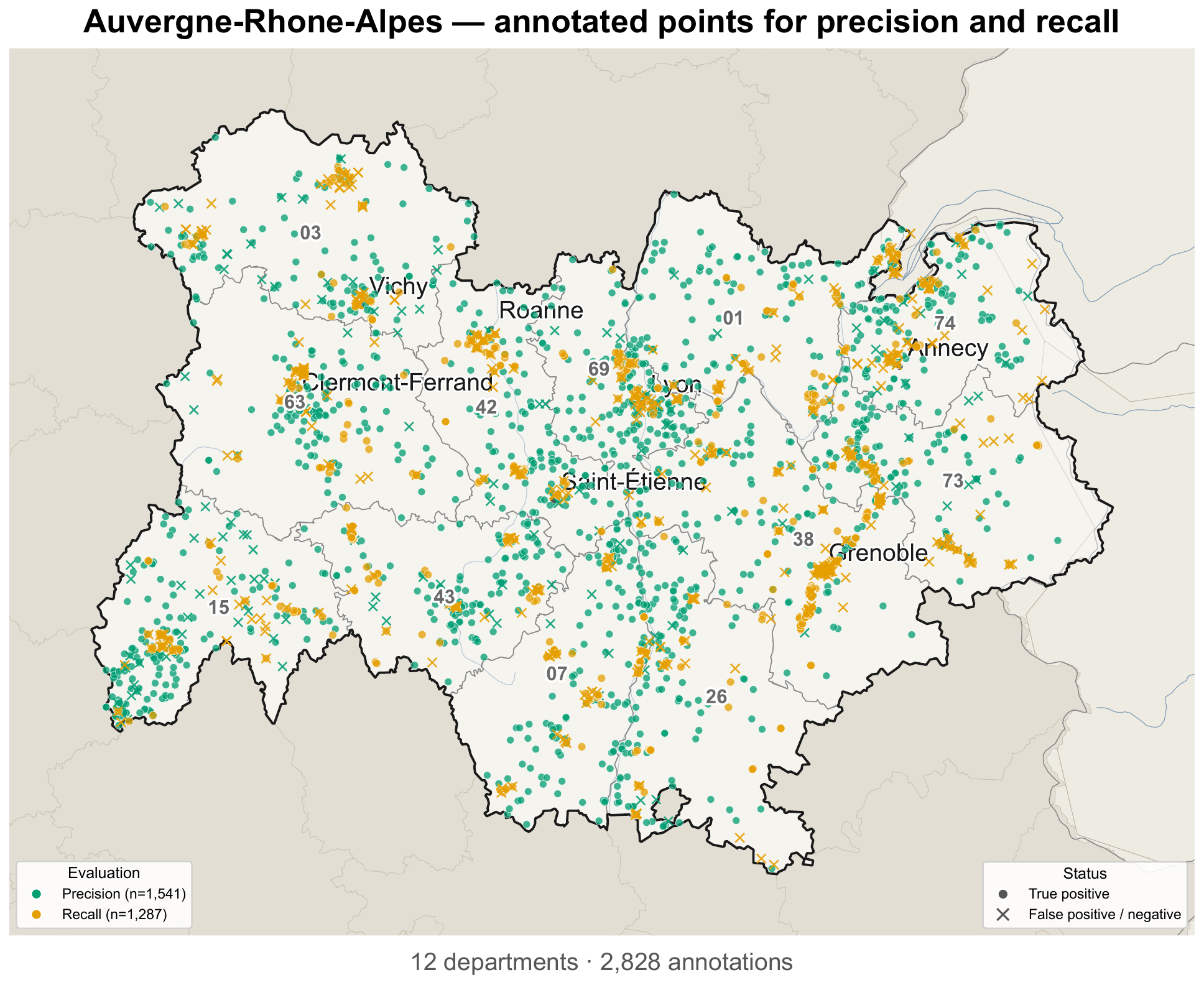}\\\noindent\includegraphics[width=0.95\textwidth]{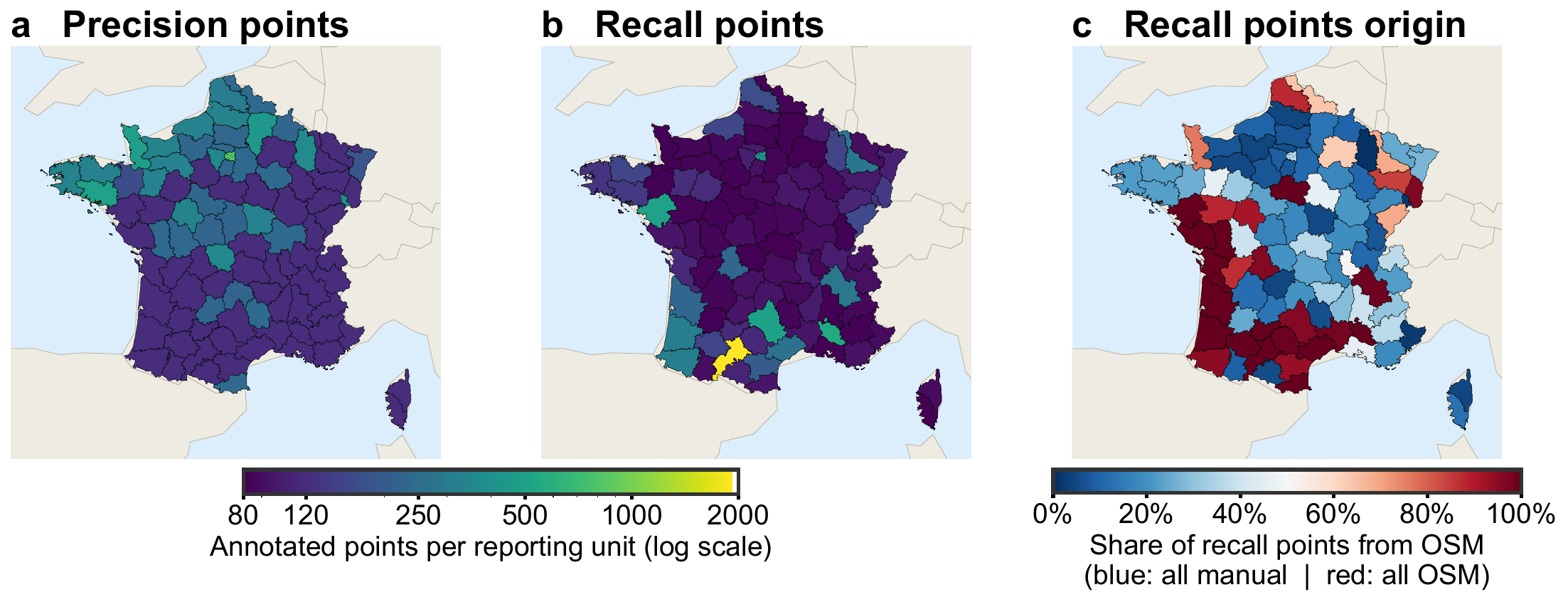}
\subsection*{Figure S3. Coverage of the precision and ground truth points.} The upper map displays the location of the sampled points over a region, green for the precision, yellow for the recall. The shape indicates whether the point was a true positive circle or a false positive (or negative) cross. The panel on the second row displays the number of precision (leftmost column), and recall (centre column) points. The rightmost plot displays the share of points coming from OSM or from a manual annotation.

\subsubsection*{Precision and recall posteriors and their uncertainty}

In practice, we used an empirical-Bayes (EB) prior for the Beta distributions of the precision and recall, rather than the uninformative Jeffreys prior. EB is the classical device for partially pooling many parallel small-sample estimates toward their common distribution\cite{robbins_empirical_1956,efron_data_1975,gelman_bayesian_2013}, fitted here on the ensemble of departments. It pools information from well-sampled units toward the smallest ones, without letting pooling manufacture or erase the underlying signal. For precision, the fitted prior weighs about 5\% of the posterior at the annotation floor; for recall, about 25\%. We checked that the choice of the prior doesn't affect the results, but only delivers tighter bounds where the number of available annotations is lower (Note S4).

Given our annotations and the true precision and recall of the raw detections, our uncertainty propagation pipeline delivers per-department half-widths (95\% CI) of $\pm$13\% where the detector performs well (F1 $\geq$ 0.7) and $\pm$21\% where it does not (F1 $<$ 0.55), for a median of $\pm$16\%. Directing extra effort at the weakest departments bounds the tail: no department exceeds $\pm$25\%. Nationally, precision reaches 0.82 and pipeline recall 0.61. These margins compare favourably with the closest exercise run without a registry to check against: an independent satellite-based estimate of Pakistan's distributed fleet, scaled nationally from a sampled territory, reports a margin of $\pm$18.2\% \cite{transitionzero_shedding_2026}. This audit reaches tighter precision for each of 93 departments individually, not for a country as a whole.

This achieved precision is wider than the $\pm$10\% back-of-the-envelope target set out above. This gap is informative rather than a shortfall in execution: the initial target was computed from the model's \emph{test-set} metrics ($P=0.81$, $R=0.96$), whereas the actual budgeting rule was applied to the pipeline's true, deployed error rates ($P=0.82$, $R=0.61$ nationally) once these were measured directly. The same 0.84-to-0.61 gap between benchmark and pipeline recall that motivates measuring pipeline-level error at all (Estimation of the precision and recall, above), as the discrepancy between the test set accuracy and the pipeline accuracy widens the achieved interval relative to the naive, test-set-based estimate.

Figure~S4 displays precision, recall, and F1 by department (top row), alongside three associated uncertainty measures (bottom row): the uncertainty on the precision estimate itself, the uncertainty on the recall estimate, and the resulting uncertainty on corrected capacity via the $P/R$ correction factor --- the same quantity plotted against F1 in Figure~S2. Precision is generally high and comparatively even across departments; recall is markedly more variable and, nationally, the weaker of the two metrics, which is visible directly in the bottom row: recall's uncertainty panel shows more spread than precision's, and it is recall, more than precision, that drives the capacity uncertainty shown in the third panel. Departments where both precision and recall are weaker in the top row correspondingly show wider capacity uncertainty in the bottom row --- the same relationship Figure~S2 establishes department by department, shown here spatially rather than as a scatter.

\noindent\includegraphics[width=0.95\textwidth]{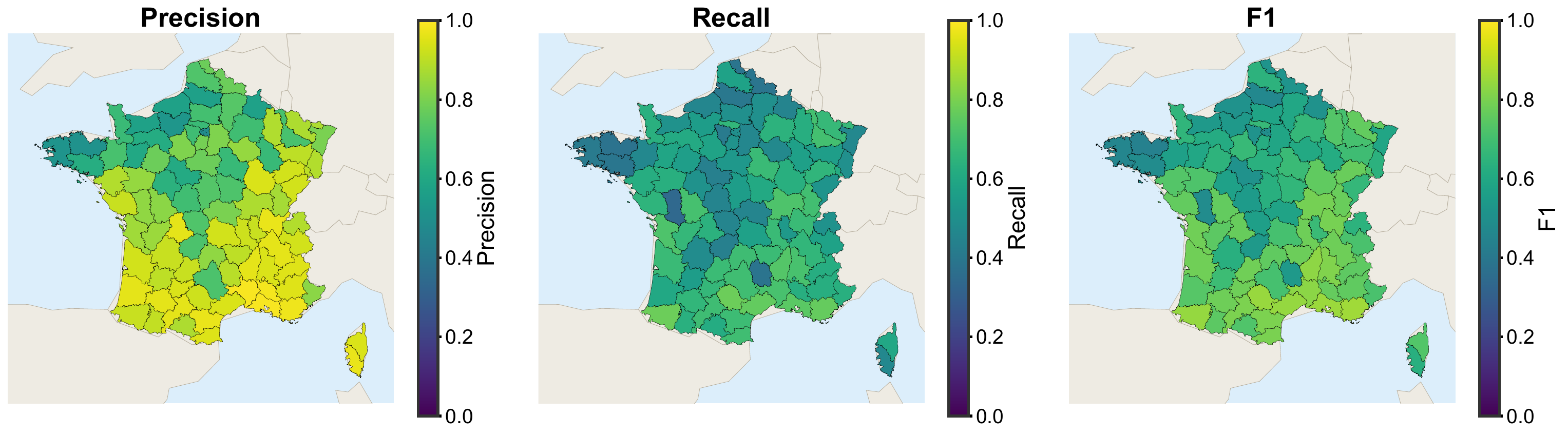}\\
\noindent\includegraphics[width=0.95\textwidth]{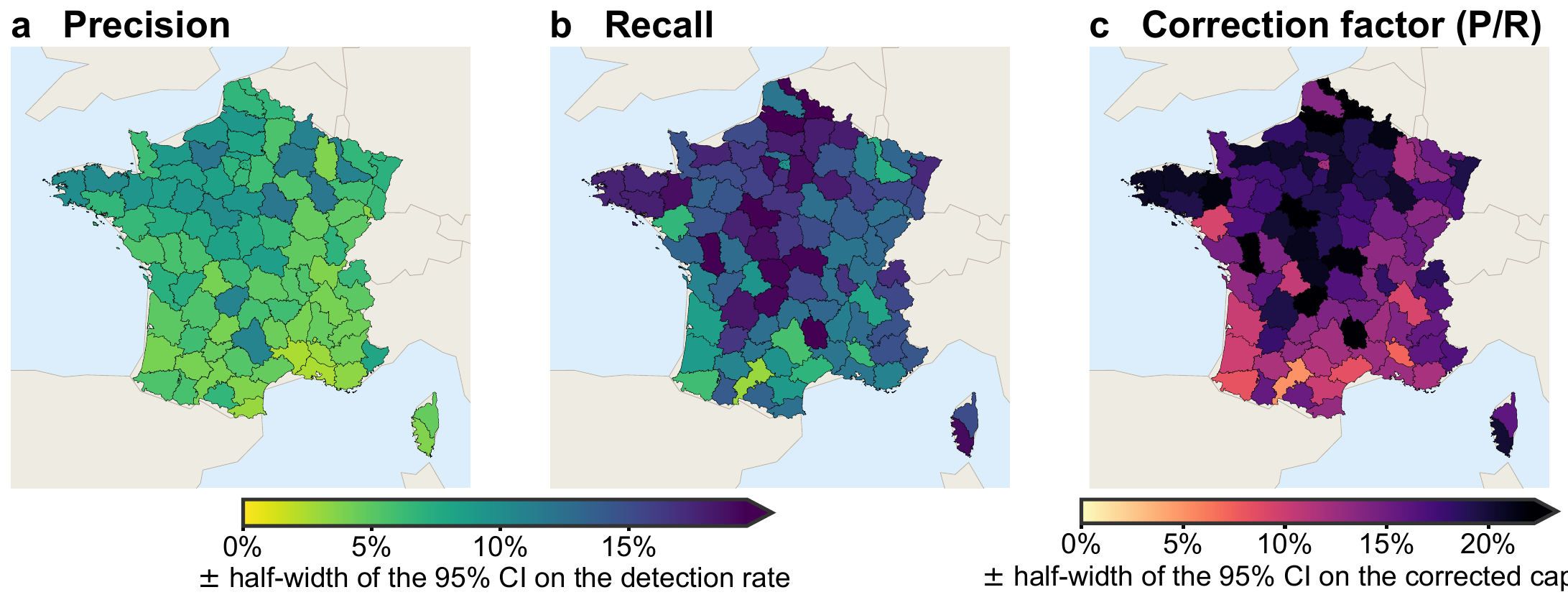}
\subsection*{Figure S4. Precision, recall, and F1 by department, with associated uncertainty.}
Top row: precision, recall, and F1 by department. Bottom row: (a) uncertainty on the precision estimate itself; (b) uncertainty on the recall estimate; (c) the resulting uncertainty on corrected capacity via the $P/R$ correction factor, the same quantity plotted against F1 in Figure~S2. Detection quality is not uniform across France. This map is descriptive of the French case: it is not a claim that heterogeneity of this kind is a general property of the method, only that it is what we observe here, and it is exactly what the department-level stratification is designed to capture.

\subsubsection*{Capacity correction}

We draw $N=10{,}000$ samples of $(P,R)$ per department and apply the correction of Component~2 at each draw --- a Monte-Carlo implementation of standard uncertainty propagation\cite{jcgm_gum_2008} --- reporting the mean and both 95\% and 99\% credible intervals from the resulting distribution. This bootstrap-based interval agrees with the closed-form prediction of Figure~S2 at a correlation of 0.999: the two are interchangeable, and the budgeting rule derived there applies here without modification.

\noindent\includegraphics[width=0.95\textwidth]{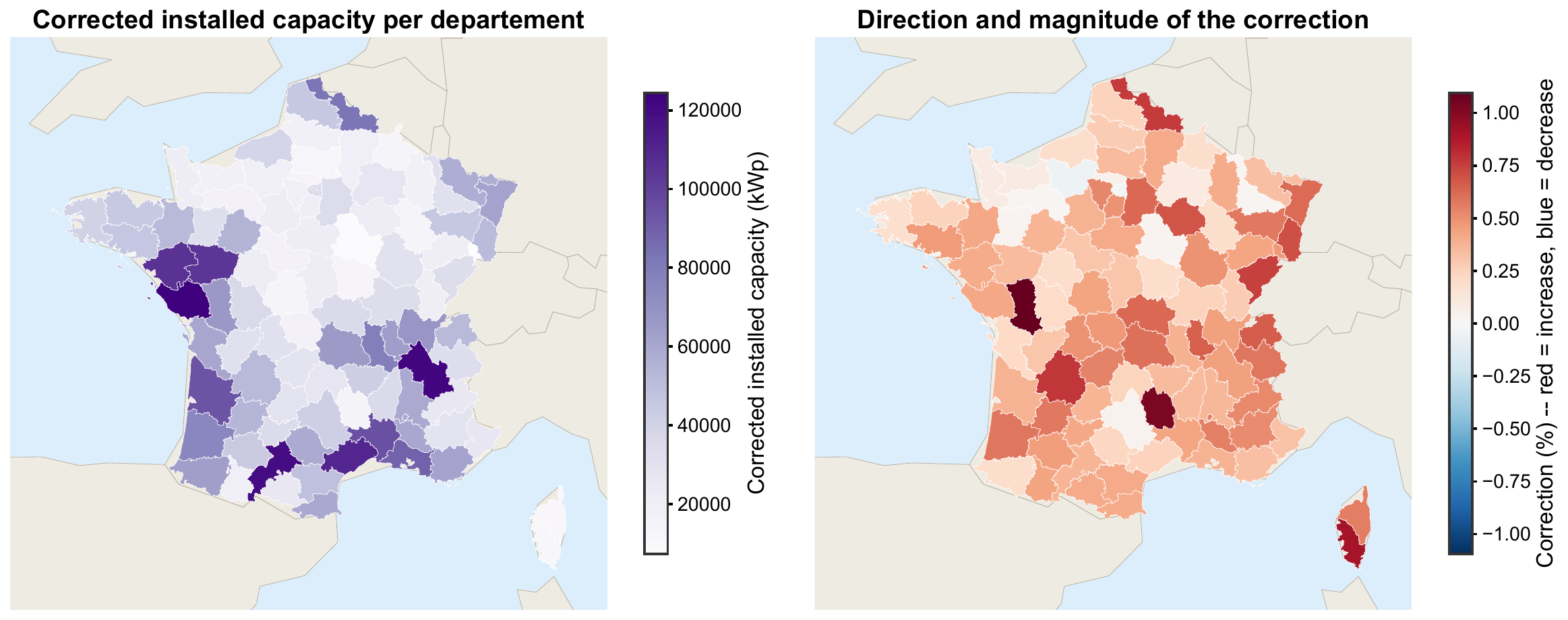}\\
\noindent\includegraphics[width=0.95\textwidth]{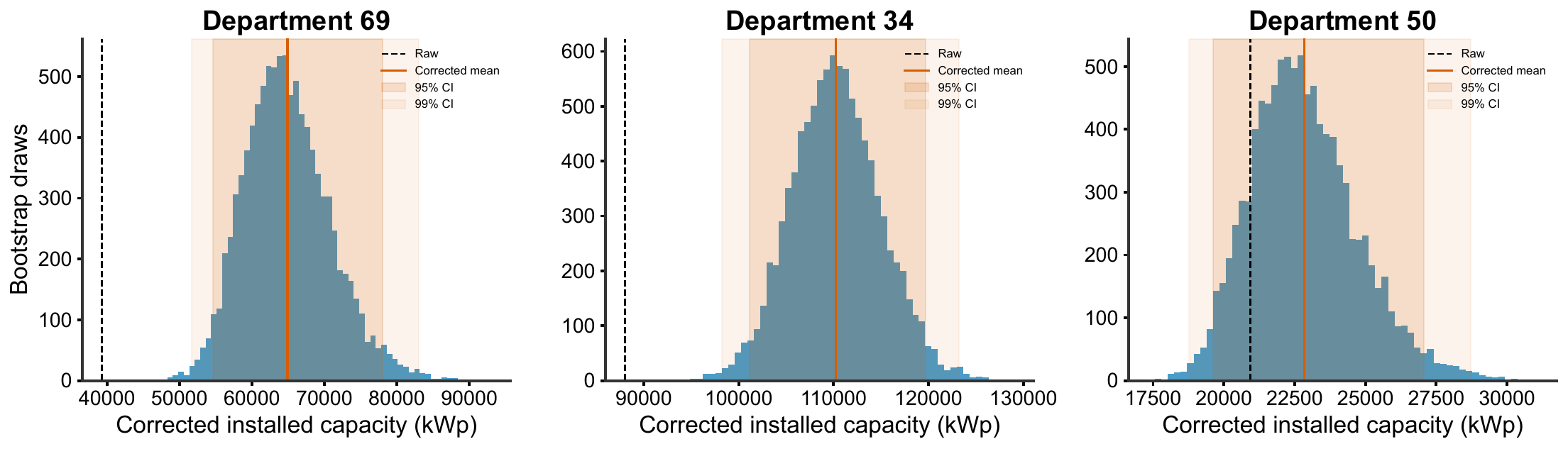}
\subsection*{Figure S5. Posterior capacity distributions.}
The upper row displays the corrected installed capacity by departement (left) and the direction and size of the correction relative to the raw detection (right). Most departements are corrected upward, consistent with the fact that the pipeline recall is low, and lower than its precision.
The lower row displays three distribution posteriors for the capacity, along with the confidence intervals plotted for each of them.

\subsubsection*{Temporal alignment: implementation}

For each municipality, the TSO reference retains only installations connected on or before the imagery acquisition date, resolved to the day. Acquisition dates are taken from the orthoimagery's own metadata; mapping campaigns are flown during the summer months. For the roughly 2,200 municipalities where the pipeline detected no installations, we instead used the acquisition date of the imagery tile covering the municipality's barycenter. We discuss the effect of coarser date granularity in Note~S7.

The public registry, the RNI, is released quarterly, but only the year-end snapshot (31 December) is archived. Since our imagery spans 2022--2025, we retrieved the four archived RNI snapshots over that period and matched each municipality to the snapshot corresponding to its acquisition year. The RNI, dated annually by construction, is matched at the year level instead of the day; because imagery acquisition falls, on average, roughly six months before the year-end cutoff, this systematically admits capacity connected after the imagery date, producing the level-artefact overshoot treated in Results and quantified fully in Note~S7.

\subsection*{A recipe for applying the protocol elsewhere}

The protocol is not tied to France, or to local units. Its main burden is the annotation effort needed to estimate the precision and recall of the raw detections, and that effort can be budgeted in advance. It depends primarily on the geographical stratification chosen for the correction, the closed-form expression of this section acting as a lower bound. In our case the minimal budget would have fallen from 19,200 samples to 2,600, a sevenfold reduction, had we reported at the regional rather than the departmental level. A single national unit would bring the requirement down to the floor itself, at the cost of returning one estimate with no spatial resolution at all.

Coarsening the grid is not free, however, and the cost is as computable as the saving. The correction is a ratio, so forming it on a coarse unit means dividing an average recall rather than averaging divided recalls. Because $1/R$ is convex, the coarse factor is always the smaller of the two, and the corrected capacity is always the lower. The penalty vanishes when recall is homogeneous within the pooled unit, whatever the dispersion of precision, and to second order it is $\mathrm{CV}(R)^2 - \rho\,\mathrm{CV}(P)\,\mathrm{CV}(R)$, evaluated within that unit. In France this amounts to 1.3\% at the regional level, roughly two thirds of the credible half-width and therefore below sampling noise (Note~S4). In a setting where recall varied three times as much between units, the same expression would put the penalty in the 5 to 9\% range.

Two consequences for anyone applying the protocol elsewhere. The direction of the bias is fixed, so a coarse audit under-corrects and is therefore conservative with respect to under-reporting: it will find less missing capacity than exists, never more. And there is no optimal grid to search for, since the finest stratification the annotation budget can support is weakly better in all cases. The variance argument that would otherwise favour coarse units is already handled by the empirical-Bayes prior, which shrinks small-sample units on precision and recall separately and therefore never subjects the ratio itself to the convexity penalty. Once the geographical stratification is chosen, the proposed approach reduces to four pieces: a posterior update per unit, and Beta priors for the precision and recall as above; the correction $C_{\text{adj}} = C_{\text{raw}}\cdot P/R$; a bootstrap over draws of $(P,R)$ to obtain a distribution per unit; and a decision rule comparing that distribution to whatever reference is under audit. Applying it elsewhere follows five steps.

\begin{enumerate}
\item \textbf{Run the detector.} Any rooftop PV detection pipeline can serve as the instrument; the audit corrects its output rather than assuming it.
\item \textbf{Define the targets and lower-bound the annotation effort.} Choose the reporting unit for stratification, and the credible-interval precision the audit needs to reach. Figure~S2's closed-form relation converts the target precision directly into a required sample size. This size should be interpreted as a lower bound, if the initial values of $P$ and $R$ are calibrated onto the test-set accuracy of the detection pipeline.
\item \textbf{Sample and annotate.} This step is the most time-consuming. The effort breaks down as:
\begin{itemize}
    \item Sample precision points, directly using the model's detections
    \item Sample recall points. For this step, crowd-sourced datasets can give a head start, supplemented by manual annotations
\end{itemize}
\item \textbf{Verify the estimation assumptions.}
\begin{itemize}
    \item {\bf H1} Ensure that the error rates are constant, or correct for heterogeneity if necessary,
    \item {\bf H2} The raw detections should cover the whole area of interest,
    \item {\bf H3} The representativeness of precision points stems from {\bf H2}. For recall points, one should ensure that the points are visible to the model (align the dates) and that their coverage is sufficient --- both geographically and in terms of coverage of the installations under scrutiny. 
    \item {\bf H4} As a general rule of thumb, capacity factors comprised between 5.0 and 6.0 can be used.
\end{itemize}
\item \textbf{Estimate the corrected capacity and compare.} Propagate the measured precision and recall through the bootstrap to obtain a calibration posterior per unit, then compare it against whatever registry is under audit --- or, where none exists, report it as the primary estimate. To facilitate the computations, we release the core source code of our analysis as a stand-alone Python package {\tt bayesian-pv-census}.\cite{kasmi_bayesian-pv-census_2026}
\end{enumerate}

\subsection*{Robustness checks}

The audit rests on two choices that a reader could reasonably contest: the coefficient used to turn detected surface into capacity, and the way detection errors are aggregated into a correction factor. This section subjects each of them to the same treatment. A choice is either anchored on an external reference, or varied across the full range of values that could plausibly have been adopted, and the flags are then recomputed. The unit-level diagnosis is stable under both operations. The hard core survives the alternative aggregation and the entire span of the conversion coefficient, and the negative control of over-reported units survives alongside it, which is what rules out an artefact of the correction procedure rather than a property of the registry. What does not survive is the national aggregate, whose sign depends on the conversion coefficient. This asymmetry is the reason the paper reports a local diagnosis and not a national correction.

\subsubsection*{Surface-to-capacity conversion} The DeepPVMapper pipeline estimates a capacity using a conversion factor fitted on crowdsourced data (see the pipeline details in Note~S8). This coefficient has a value of 6.3~m$^2$/kWp, which, once corrected for tilt to apply to planimetric area, corresponds to a value of 5.5~m$^2$/kWp. This value is a single national average, applied to a fleet that has been installed over the course of twenty years. Besides, BDPV is skewed towards older systems, so lower efficiency values may inflate the area required per kWp.

To test whether the conversion coefficient could yield detection artifacts, we anchored it on the registry itself, on a subset chosen to avoid circularity. The anchor is computed only on the 60 units whose registry value already falls inside the credible interval of the corrected estimate. On this subset the two size distributions should coincide: on the one hand the distribution of capacities estimated with DeepPVMapper and on the other hand, the distribution of capacities registered in the TSO connection data. Note that the convergence is on the distribution, not the total installed capacity. This design avoids a potential circularity in our evaluation of the methodology. 

These units are, by construction, not the ones whose status the audit is trying to establish, so the coefficient is not calibrated on the discrepancy it is later used to measure. On this subset the two size distributions should coincide. The median detected array carries 3.07~kWp against 3.00~kWp for the median registry record, so the coefficient implied by the registry is $5.5 \times 3.07/3.00 = 5.63$~m$^2$/kWp, within 2.4\% of the retained value. Quantile by quantile, the converted distribution reproduces the registry distribution to within 5\% across the interquartile range. The upper quantiles diverge, which is expected: the registry side is shaped by the 3, 6 and 9~kVA contractual tiers, and the two sides do not count the same object, since extensions and contract changes split a system into several records on the registry side while clustering merges them into one array on the detection side. 

This test enables us to reject the hypothesis that the conversion factor used in DeepPVMapper carries quantifiable biases in the estimation of the rooftop PV capacity in France. In addition to this result, we study the sensitivity of the results of a variation of the conversion coefficient from 5.0 to 6.0~m$^2$/kWp as an item of the specification battery discussed in the next section. Figure S6 displays the comparison of the distributions and the Q-Q plot between the estimated and registered installed capacities. Two features stand out. The registry distribution is discretised: 26.6\% of its records sit at exactly 3.0~kWp and a further 13.3\% at 6.0~kWp, so its quantile function is a staircase, and a quantile-by-quantile comparison measures that staircase rather than the coefficient. Read on central tendency instead, the two distributions agree: the implied coefficient is 5.63~m$^2$/kWp on medians and 5.71~m$^2$/kWp on means, both inside the specification range. Both also sit slightly above the retained 5.5, so the pipeline is, if anything, marginally generous in capacity per square metre. That is the direction that works against the under-reporting result, not towards it.

\noindent\includegraphics[width=0.95\textwidth]{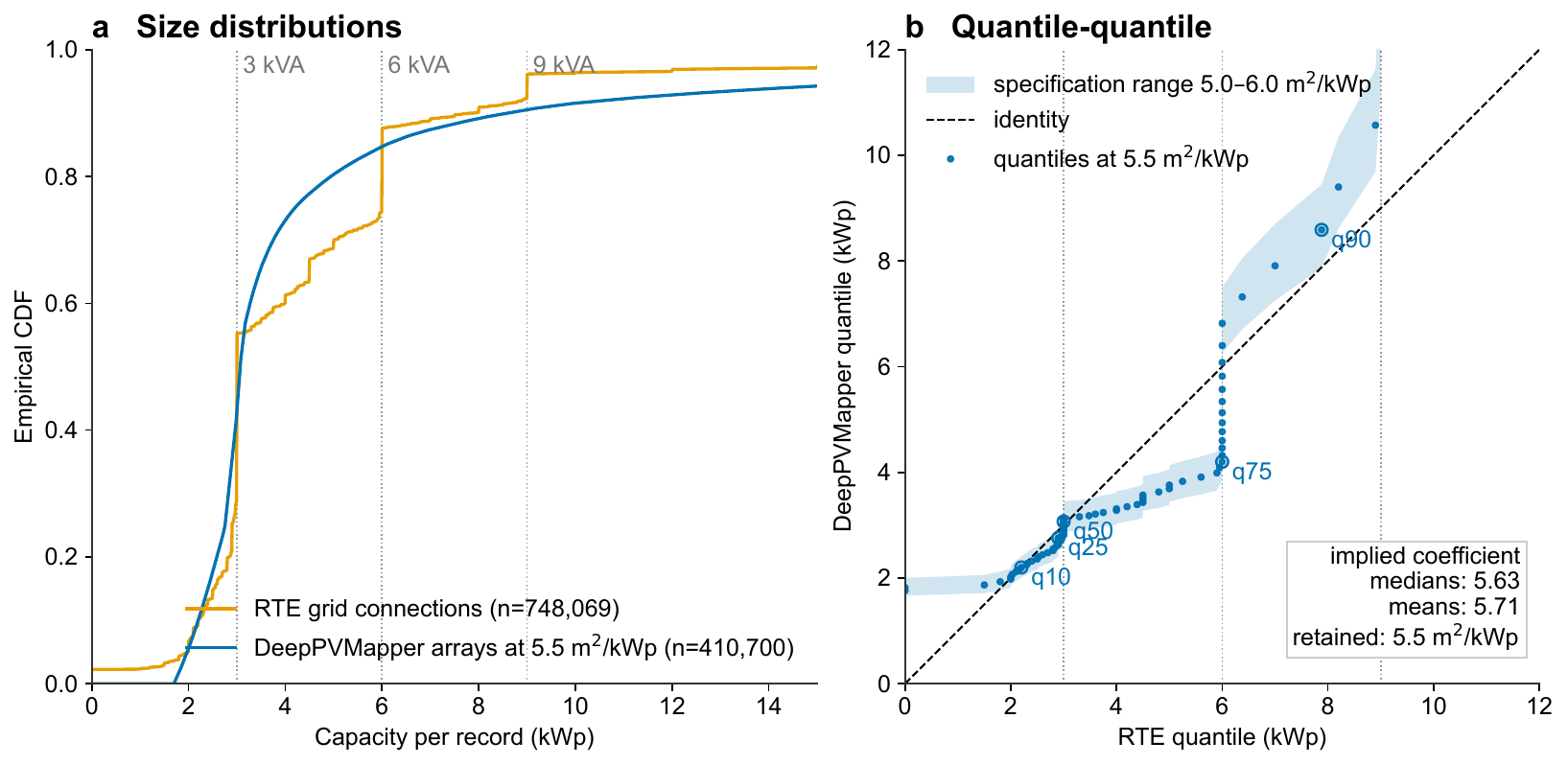}
\subsection*{Figure S6. Surface-to-capacity agreement.}
Both panels use the 60 reporting units whose registry value already falls inside the credible interval of the corrected estimate, the non-circular subset defined above (748,069 grid connections, 410,700 detected arrays). (a) Empirical cumulative distribution of capacity per record: RTE grid connections (orange) and DeepPVMapper arrays converted at 5.5~m$^2$/kWp (blue). Dotted lines mark the 3, 6 and 9~kVA contractual tiers, on which the registry places 26.6\%, 13.3\% and 3.9\% of its records. (b) Quantile-quantile plot of the same two distributions, identity line dashed, selected quantiles circled and labelled. The shaded band is the specification range: the interval spanned by the detected quantiles when the conversion coefficient is moved to its bounds, 5.0 and 6.0~m$^2$/kWp. Vertical runs of points occur where the registry quantile is pinned to a contractual tier while the detected quantile keeps rising.

\subsubsection*{Specification batteries.} 

\paragraph{Sensitivity to recalibration factor}

A unit is flagged as under- (over-) reported when the registry value falls below (above) the 99\% credible interval of its corrected bootstrap distribution. This distribution is centered around the unit's true installed capacity provided that assumptions {\bf H1}--{\bf H4} hold. As discussed in the Methods (Practical implementation for the French case study), empirical tests of {\bf H1} indicated a limited dependency on the precision to the system size. 

Therefore, provided an alternative definition of Equation (3), where instead of reweighting the estimated capacity by $P/R$, we instead re-weight it by $P_w/R_w$, taking into account the slight precision and recall variations across the system sizes.

The baseline variant of our methodology is described as the Specification {\bf A}. The second specification, which incorporates a size-sensitive re-weighting factor $P_w/R_w$ is referred to as the Specification {\bf B}. Note that this specification {\bf B} does not assume H1 and can therefore be used to carry out our methodology over a full fleet of PV systems ({\bf B} falls back to count-based recall wherever fewer than 40 surface-bearing ground-truth points are available in a unit — a contingency built into the specification but not triggered in practice for any of the 93 units). The decision rule (flag a unit if its recorded value lands outside the 99\% credible interval) remains unchanged. For Specification {\bf B}, the conversion coefficient also remains the same, at 5.5~m$^2$/kWp.

We can define the \textbf{hard core} as the set of units flagged unanimously by \textbf{A} and \textbf{B}. A unit flagged by \textbf{A} but not by \textbf{B} is not a unit where the discrepancy vanishes: it is a unit whose discrepancy survives the point estimate but not the wider interval that size-dependent detection errors produce, and it is therefore reported as a weaker case rather than as an absence. Figure~1 in the Results section displays the unit flagged and the hard core set of units. The units flagged by Specification {\bf A} are the following. Departments in italics are shared by both specifications, and therefore constitute the hard core of flagged units. 

\begin{itemize}
    \item \underline{Under-report:} \emph{Bouches-du-Rhône} (13), \emph{Calvados} (14),
    \emph{Cantal} (15), \emph{Corrèze} (19), \emph{Corse-du-Sud} (2A), \emph{Creuse} (23),
    \emph{Deux-Sèvres} (79), \emph{Dordogne} (24), \emph{Eure-et-Loir} (28),
    \emph{Haute-Corse} (2B), \emph{Haute-Saône} (70), Haute-Vienne (87), Indre (36),
    Loir-et-Cher (41), Lot-et-Garonne (47), \emph{Lozère} (48), Maine-et-Loire (49),
    \emph{Meuse} (55), Nièvre (58), \emph{Orne} (61), \emph{Territoire de Belfort} (90),
    Val-d'Oise (95), \emph{Vienne} (86), \emph{Vosges} (88), and \emph{Paris and its inner ring}
    (75, 92, 93, 94).
    \item \underline{Over-report:} \emph{Eure} (27), \emph{Haute-Garonne} (31), Hérault (34),
    \emph{Indre-et-Loire} (37), \emph{Manche} (50), \emph{Meurthe-et-Moselle} (54),
    \emph{Tarn} (81), \emph{Yonne} (89).
\end{itemize}
The units flagged by Specification {\bf B} are the following, with the same italics convention.
\begin{itemize}
    \item \underline{Under-report:} Alpes-de-Haute-Provence (04), Bas-Rhin (67),
    \emph{Bouches-du-Rhône} (13), \emph{Calvados} (14), \emph{Cantal} (15), \emph{Corrèze} (19),
    \emph{Corse-du-Sud} (2A), \emph{Creuse} (23), \emph{Deux-Sèvres} (79), \emph{Dordogne} (24),
    Doubs (25), \emph{Eure-et-Loir} (28), \emph{Haute-Corse} (2B), \emph{Haute-Saône} (70),
    \emph{Lozère} (48), \emph{Meuse} (55), \emph{Orne} (61), \emph{Territoire de Belfort} (90),
    Vendée (85), \emph{Vienne} (86), \emph{Vosges} (88), and \emph{Paris and its inner ring}
    (75, 92, 93, 94).
    \item \underline{Over-report:} \emph{Eure} (27), Gironde (33), \emph{Haute-Garonne} (31),
    \emph{Indre-et-Loire} (37), Landes (40), \emph{Manche} (50), Marne (51), Mayenne (53),
    \emph{Meurthe-et-Moselle} (54), Pyrénées-Orientales (66), Savoie (73), \emph{Tarn} (81),
    Var (83), \emph{Yonne} (89).
\end{itemize}

Finally, Table~S1 presents the concordance table between the two specifications.

\subsection*{Table S1. Concordance of the two specifications over the 93 reporting units} 
The diagonal cells in bold define the hard core: 18 units flagged under-reported by both specifications, and 7 flagged over-reported by both, the negative control. The two off-diagonal zeros are the substantive result: no unit changes sign between specifications.

\begin{tabular}{lccc}
\toprule
 & \multicolumn{3}{c}{Specification B ($P_w/R_w$)} \\
\cmidrule(lr){2-4}
Specification A ($P/R$) & below & within & above \\
\midrule
below   & \textbf{18} & 7 & \textbf{0} \\
within  & 4 & 49 & 7 \\
above   & \textbf{0} & 1 & \textbf{7} \\
\bottomrule
\end{tabular}

\paragraph{Sensitivity to the conversion coefficient}

Additionally, we investigated the sensitivity of our findings to a variation in the conversion coefficient. In the previous section we showed that our conversion factor is well-calibrated against the connection data. As a robustness check, we studied the impact of varying the conversion coefficient from 5.0 to 6.0~m$^2$/kWp to see how this affects our findings. We refer to this specification as the Specification {\bf C}. Figure S7 presents the effect of the conversion factor on the flag status for underreporting and overreporting units. 

\noindent\includegraphics[width=0.95\textwidth]{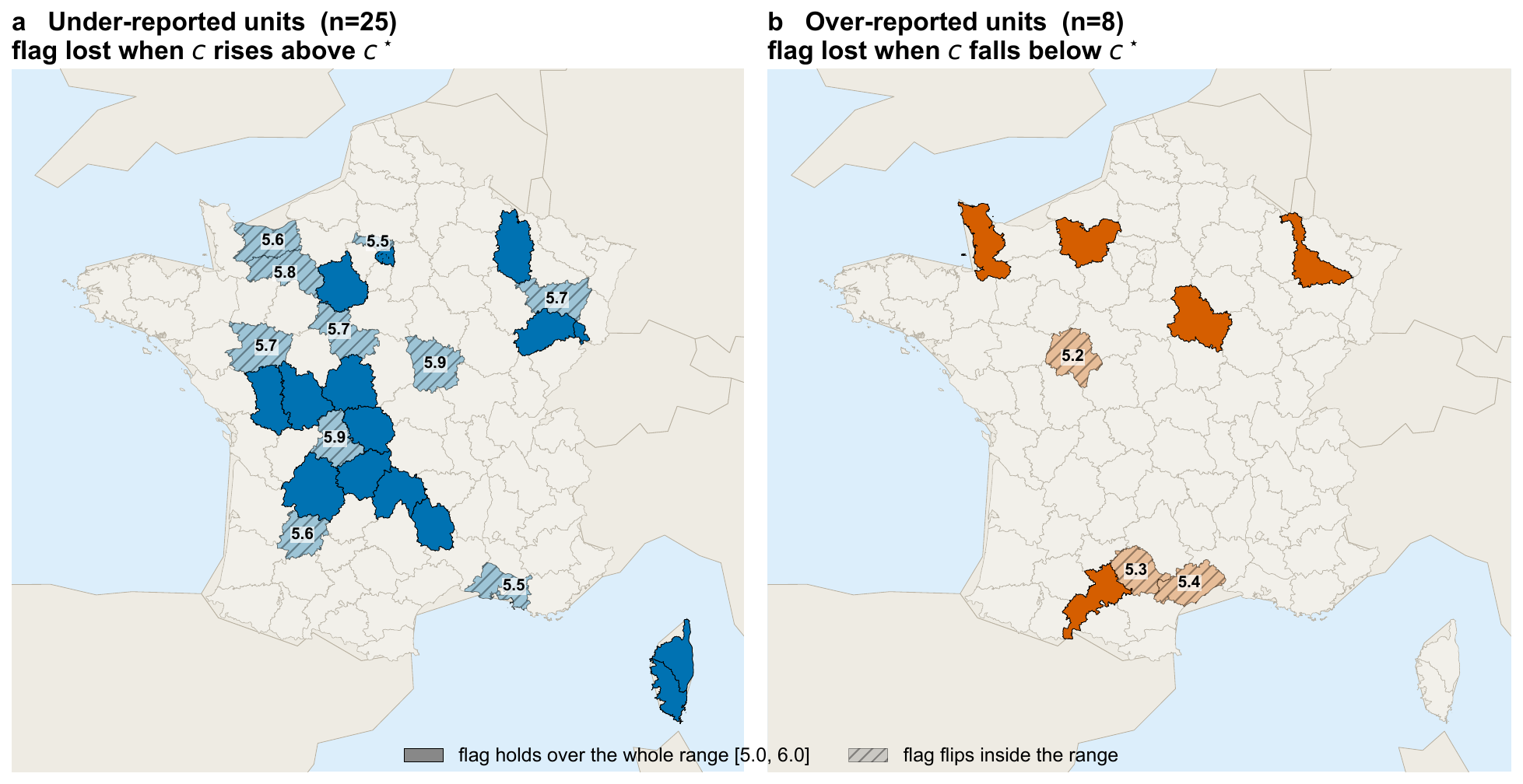}
\subsection*{Figure S7. Effect of the conversion coefficient on the flag status.}
A unit's status is monotone in the conversion coefficient and flips at a single threshold $c^\star$, available in closed form as $c^\star = 5.5 \times \text{CI}_{99}/\text{registry}$, with the lower bound for under-reported units and the upper bound for over-reported ones. No numerical sweep is needed, and the closed form reproduces exactly the statuses recomputed at 5.0 and 6.0~m$^2$/kWp. Only the units flagged at the baseline coefficient are shown: (a) the 25 under-reported units, (b) the 8 over-reported units. The number printed on each unit is its $c^\star$. The two directions are mirror images: an under-reported unit loses its flag when the coefficient \emph{rises} above $c^\star$, an over-reported unit when it \emph{falls} below. Solid fill marks the units whose flag holds across the whole specification range 5.0--6.0~m$^2$/kWp, 15 of 25 under-reported and 5 of 8 over-reported; hatching marks those that flip inside the range. Units classified as \emph{within} at the baseline are not shown: 12 of them acquire an under-report flag and 11 an over-report flag somewhere in the range, but colouring them would fill the map without adding information.

\subsubsection*{Empirical strategy of the fragmentation mechanism.}

\paragraph{Empirical model}

The distributor serving each municipality is read from the operator field of the connection data and aggregated to the municipal level. We exclude two sets of municipalities: 204 served by more than one operator, and 198 carrying no operator label at all. The latter arise because the label is derived from the registry itself, so a municipality entirely absent from it has no operator to read. This second exclusion is inconsequential for the intensive margin, which conditions on positive registry capacity and where the label is therefore always available. It does bear on the extensive margin: among these 198 municipalities, 28\% have zero registry capacity, against 0.35\% in the estimation sample --- so the extensive-margin estimate is a lower bound, one further reason to read it as directional rather than established.

We intersect this sample with municipalities where DeepPVMapper detects at least 50~kWp of rooftop PV, which leaves $n = 14{,}551$ municipalities, 817 of them ELD-served. Corsica's island operator is treated as an ELD. Below that floor the coverage ratio is dominated by detection noise on a handful of roofs. The estimating equation regresses the log ratio of registry to detected capacity on an ELD indicator, with department fixed effects:

\begin{equation}
\log\!\left(\frac{C^{\text{reg}}_{i}}{C^{\text{det}}_{i}}\right)
= \beta\,\text{ELD}_{i} \;+\; \gamma' X_{i} \;+\; \delta_{d(i)} \;+\; \varepsilon_{i},
\label{eq:eld}
\end{equation}
where $C^{\text{reg}}_{i}$ and $C^{\text{det}}_{i}$ are the registry and detected capacities of municipality $i$, $\text{ELD}_{i}$ indicates that $i$ is served by a local distribution company, $X_{i}$ collects log detected capacity and log population, and $\delta_{d(i)}$ is a fixed effect for the department containing $i$. The quantity of interest is $\beta$, and $\exp(\beta)-1$ is the proportional gap in registry coverage between the two groups of the same department.

Two properties follow from the design. The recalibration factor is departmental, therefore constant within a fixed effect, so it cancels exactly and no element of the calibration enters the coefficient. Imagery year is likewise constant within every department, so a differential imagery date between ELD-served and incumbent-served municipalities is absorbed by construction rather than assumed away. This is why the test is conducted at the municipal level and not at the level of the reporting unit.

\paragraph{Identification assumption.} 

We assume that, conditional on department and controls, ELD status is uncorrelated with any other determinant of the registry-to-detection ratio. What makes this plausible is institutional rather than statistical. ELD perimeters are a legacy of the 1946 nationalisation, which exempted pre-existing municipal utilities and cooperatives from transfer to the national operator. The assignment is eighty years old and unrelated to PV deployment in the 2020s, which rules out reverse causation. It does not amount to random assignment. The coefficient is therefore reported throughout as a conditional association, not as a causal effect.

The leading candidate mechanism is institutional rather than behavioural: RTE and the incumbent DSO share a common lineage and decades of joint reporting infrastructure inherited from their unbundling out of a single vertically-integrated utility, whereas each ELD independently built and maintains its own interface to RTE's information system. Consistent with an integration-based account rather than a reporting-frequency one, the gap does not track the regulatory reporting-frequency threshold alone (Results).

The assumption bears on unobservables and cannot be tested directly. What can be shown is balance on observables, as indirect evidence: if the two groups already diverged on what we measure, their comparability on what we do not measure would be doubtful. Table~S2 reports it. ELD-served municipalities are smaller on average, which is why size enters the specification as a control, and the imbalance is mild once expressed in logs, the form in which the covariates enter
Equation~8.

\subsection*{Table S2. Covariate balance between ELD-served and incumbent-served municipalities.} Sample as in Equation 8: 14,551 municipalities with at least 50~kWp of detected PV, 817 of them ELD-served. The normalised difference is the difference in group means divided by the pooled standard deviation; values beyond 0.25 in absolute value are conventionally taken to indicate problematic imbalance. The within-department difference is the coefficient on the ELD indicator in a regression of the covariate on department fixed effects, with cluster-robust standard errors in parentheses and wild cluster bootstrap $p$-values (299 replications).

\begin{tabular}{lccccc}
\toprule
 & \multicolumn{2}{c}{Group mean} & Normalised & \multicolumn{2}{c}{Within-department} \\
\cmidrule(lr){2-3}\cmidrule(lr){5-6}
 & ELD & Incumbent & difference & difference & $p$ (wild) \\
\midrule
Population              & 2,200 & 3,907 & $-0.102$ & $-1{,}230$ & 0.040 \\
                        &       &       &          & (563) &  \\
Detected capacity (kWp) & 150.8 & 187.7 & $-0.161$ & $-27.5$ & 0.023 \\
                        &       &       &          & (12.2) &  \\
Detected arrays         & 24.2  & 34.9  & $-0.255$ & $-7.42$ & 0.033 \\
                        &       &       &          & (3.30) &  \\
log population          & 7.042 & 7.170 & $-0.108$ & $-0.244$ & 0.210 \\
                        &       &       &          & (0.172) &  \\
log detected capacity   & 4.769 & 4.873 & $-0.150$ & $-0.074$ & 0.087 \\
                        &       &       &          & (0.041) &  \\
\bottomrule
\end{tabular}

\paragraph{Threat and falsification.} The principal threat is that ELD municipalities are systematically less well detected. That would depress the denominator and inflate the apparent gap. Column (5) of Table~S3 tests this directly by regressing detected capacity alone on the ELD indicator: the two groups do not differ within a department. The bias would in any case have run against the result, since an understated denominator raises apparent coverage rather than lowering it. 

\paragraph{Inference.} Standard errors are clustered by department. Only 32 of the 96 departments contain both operator types and therefore carry identifying variation, and at that number the asymptotic cluster-robust standard error is biased downwards. We report wild cluster bootstrap $p$-values with Rademacher weights and the null imposed, 999 replications~\cite{cameron2008bootstrap}. The extensive margin is estimated as a linear probability model, a fixed-effects logit being subject to the incidental parameters problem with 96 groups. Table~S3 presents the regression results. 

\subsection*{Table S3. Registry coverage in ELD-served versus incumbent-served municipalities of the same department.} 
Cluster-robust standard errors in parentheses, wild cluster bootstrap $p$-values in brackets (999 replications, Rademacher weights, null imposed). Controls are log detected capacity and log population; imagery year is constant within department and therefore absorbed by the fixed effects. Columns (1) and (2) estimate Equation 8 on municipalities with positive registry capacity, without and with controls; the implied gaps are $-42\%$ and $-38\%$. Columns (3) and (4) estimate the extensive margin as a linear probability model: $+3.6$ percentage points against an incumbent baseline of $0.24\%$. That margin rests on 18 ELD municipalities with no registry record and is not statistically resolved, so it is reported as directionally consistent only. Column (5) is the falsification: detected capacity alone does not differ between the two groups, which rules out a detection artefact. Standard errors are clustered by department (the 96 administrative departments, prior to the Paris–inner-ring merger used elsewhere in this audit for annotation purposes) of which 32 contain both operator types and therefore carry the identifying variation.
\begin{tabular}{lccccc}
\toprule
 & \multicolumn{2}{c}{log registry coverage} & \multicolumn{2}{c}{no registry record} & falsification \\
\cmidrule(lr){2-3}\cmidrule(lr){4-5}\cmidrule(lr){6-6}
 & (1) & (2) & (3) & (4) & (5) \\
\midrule
ELD municipality & $-0.538$ & $-0.482$ & 0.036 & 0.036 & 0.029 \\
 & (0.133) & (0.162) & (0.020) & (0.019) & (0.053) \\
 & [0.001] & [0.011] & [0.123] & [0.123] & [0.607] \\
\midrule
Department fixed effects & Yes & Yes & Yes & Yes & Yes \\
Controls & No & Yes & No & Yes & Yes \\
Observations & 14,500 & 14,500 & 14,551 & 14,551 & 14,551 \\
Clusters & 96 & 96 & 96 & 96 & 96 \\
\quad of which identifying & 32 & 32 & 32 & 32 & 32 \\
Within $R^2$ & 0.019 & 0.273 & 0.009 & 0.011 & 0.487 \\
\bottomrule
\end{tabular}

\paragraph{Non-parametric corroboration.} The regression assumes a log-linear form and pools all municipalities. We therefore repeat the comparison without either assumption. Within each of the 19 departments where both operator types have at least five municipalities, we take the median coverage of each group and test the 19 paired differences with a Wilcoxon signed-rank test. The difference is negative in 16 of the 19 departments ($p = 0.0003$; Figure~2B). This says nothing about magnitude, which is the price of dropping the functional form, but it establishes the direction without relying on it.

\paragraph{Robustness.} The 50~kWp floor is a choice of the analyst, and it conditions on the denominator and therefore on its measurement error. We re-run the entire procedure at 20 and 100~kWp. The estimate of $\beta$ lies between $-0.45$ and $-0.54$ without controls, and between $-0.40$ and $-0.48$ with them --- a proportional gap of $-36$\% to $-42$\% without controls ($-33$\% to $-38$\% with them). The sign never changes. The Wilcoxon test rejects at all three floors, on a number of paired departments that falls from 22 to 9 as the floor rises. The extensive margin is resolved at none of them, which confirms that it should be read as directional only. One caveat is worth stating plainly: with controls at the 100~kWp floor, $p = 0.07$. The sample is then almost three times smaller than at 20~kWp while the point estimate barely moves, so this is a loss of power and not a reversal.

\subsection*{Data sources}

\paragraph{TSO connection data.} We had access to the French TSO (RTE) connection data. This dataset contains all assets connected within RTE's control area. It is arguably the ground truth dataset regarding the French electric system. Each installation connected to the French grid is reference by an internal identification number, its connection date and its connected capacity. Additional information regarding the lifecycle of a power generation unit (upgrade, removal) is also referenced and traced back. Therefore, the TSO connection data does not only contains the current view of the electric system, but also all its past history. 

For this study, we focused on a subset of the TSO's registry, containing only rooftop-related PV assets. For each connected PV system, we retrieve the connection date, whether it was still active at the image acquisition date, the installed capacity and the city where the system is located. The city information enables us to geocode the installations at the city and departemental levels. We filter out all PV systems that have a capacity greater than 36~kWp. 

The audited reference (before temporal alignment with the image acquisition date) comprises 1,034,283 installations totalling 4.89~GWp. Taking into account the temporal alignment with the orthoimagery, the capacity drops to 3.90~GWp and the number of installations to 853,469.

\paragraph{Public registry (RNI).} The {\it Registre national d'installations} (RNI) is the French open data source for grid connections. It is maintained, curated and published by RTE. The RNI is released quarterly and only the final yearly publication (at the 31 December) is archived. Contrary to the connection data, PV systems with an installed capacity lower than 36~kWp are aggregated at the city level. Besides, when the records in a given municipality comprise less than 10 systems, the capacity is not disclosed, following a censoring rule set by the law~\cite{republique_francaise_article_2016}. The RNI is accessible on the public platform {\it Open Data Réseaux Energie} (ODRE). 

The capacity that is not published at the city level is not lost: it is simply aggregated at the departmental level. We build two audit variants: \emph{as published} at municipal level (truncated), and \emph{truncation-corrected}, reintegrating the aggregated capacity at the departmental level. 

\paragraph{Aerial orthoimagery.} Raw detections are produced on national aerial orthoimagery distributed by the French national mapping agency (IGN)\cite{ign_bd_2024}, at a native resolution of 20 cm/pixel. Coverage is national but not simultaneous: individual d\'epartements are re-flown on a rotating multi-year cycle, primarily during the summer months, so imagery vintage varies by d\'epartement within the study window. For this study, imagery spans 2022--2025 (see Note~S6); each municipality is matched to the acquisition date of its own most recent available tile, both for detection and for the day-level temporal alignment described in \emph{Component 3}. Imagery is accessed via the public portal \url{https://cartes.gouv.fr/rechercher-une-donnee/dataset/IGNF_BD-ORTHO}.

\section*{RESOURCE AVAILABILITY}

\subsection*{Lead contact}
Requests for further information and resources should be directed to and will be fulfilled by the lead
contact, Gabriel Kasmi (gabriel.kasmi@minesparis.psl.eu).

\subsection*{Materials availability}
This study did not generate new materials.

\subsection*{Data and code availability}

\begin{itemize}

\item \textbf{Detection product.} The rooftop PV detections analysed in this study correspond to the file \texttt{latest\_dpvm.geojson} of release v3 of the OpenPVMapper database\cite{kasmi_openpvmapper_2026}, deposited at Zenodo: \url{https://doi.org/10.5281/zenodo.21534856} and publicly available as of the date of publication.

\item \textbf{Replication data.} All data needed required to reproduce the results, tables and figures of this study have been deposited at Zenodo: \url{https://doi.org/10.5281/zenodo.22729786}, and are publicly available as of the date of publication. All but two results are reproducible (due to disclosure contraints). See the below the commentary on reproducibility.

\item \textbf{Third-party sources.} The public registry (RNI), the BD ORTHO orthoimagery and its mosaicking graphs and the OpenStreetMap extract are openly available from their respective providers, as described in the Methods (\emph{Data sources}). 

\item \textbf{Code.} All original code is open source and released under an MIT license. Several code assets are made available upon publication of this work:
\begin{itemize}
    \item The statistical core of the audit, which turns raw detections into a posterior over installed capacity and evaluates are ported value against it, is distributed as a standalone Python package, \texttt{bayesian-pv-census}\cite{kasmi_bayesian-pv-census_2026}. Its source is at \url{https://github.com/gabrielkasmi/bayesian-pv-census} and it is archived at Zenodo: \url{https://doi.org/10.5281/zenodo.21921771}. 
    \item The full analysis code of the study, from the raw detections to every figure, is available at \url{https://github.com/gabrielkasmi/pv-registry-audit} and archived at Zenodo:\url{https://doi.org/10.5281/zenodo.21925115}.
    \item The annotation tooling used to build the validation samples can be accessed at \url{https://github.com/gabrielkasmi/pv-annotation}.
    \item The source code of the mapping algorithm DeepPVMapper is accessible at \url{https://github.com/gabrielkasmi/deeppvmapper}
\end{itemize}

\item \textbf{Commentary on reproducibility.} The raw connection data extract cannot be redistributed. The departmental aggregates released with this paper contain every quantity used in the audit, and are released with the operator's authorisation. Municipality-level registry values are not released, since at that granularity they would disclose what the public registry withholds under its own privacy rule. This has three implications regarding the reproducibility of the results. First, the municipality-level date matching is provided but cannot be reproduced. Second, the municipality-level fixed-effects regression of the distribution-chain test cannot be reproduced, whereas its non-parametric counterpart can. The departmental medians by operator type that we release reproduce it exactly, with 16 of 19 departments negative and a Wilcoxon signed-rank $p = 0.0003$. The mechanism finding therefore rests on a test that any reader can re-run. Third, the external anchoring of the surface-to-capacity coefficient pools the capacity distribution of every consistent unit, which the departmental aggregates do not carry. We release capacity deciles for three illustrative units, which reproduce the congruence between the two size distributions and place the implied coefficient inside the 5.0--6.0~m$^2$/kWp range, without recovering the anchoring value itself. 

\item \textbf{Additional information.} The intent of this release is that the audit be reproducible end to end without contacting the authors. Any additional information required to reanalyse the data reported in this paper is nevertheless available from the lead contact upon request.

\end{itemize}

\section*{ACKNOWLEDGMENTS}
The authors thank Augustin Touron and the teams at DIA\textsuperscript{2} (RTE) for their help in accessing and interpreting the grid connection data, and for sharing their operational knowledge of how the registry is built and maintained. This audit would not have been possible without that valuable feedback.

\section*{AUTHOR CONTRIBUTIONS}
Conceptualization, G.K.; methodology, G.K.; software, G.K.; formal analysis, G.K.; investigation, G.K.;
data curation, G.K.; visualization, G.K.; writing---original draft, G.K.; writing---review \& editing,
G.K., Y.-M.S.-D., L.D., and P.B.; supervision, Y.-M.S.-D., L.D., and P.B.; funding acquisition, L.D. and
P.B.

\section*{DECLARATION OF GENERATIVE AI AND AI-ASSISTED TECHNOLOGIES}

During the preparation of this work, the author(s) used Anthropic Claude Cowork in order to help with the creation of the source code and notebooks that underpin the analysis, adjust the layout of the figures and tables and help with the redaction of the final manuscript. After using this tool or service, the author(s) reviewed and edited the content as needed and take(s) full responsibility for the content of the publication.

\section*{DECLARATION OF INTERESTS}
G.K.\ and L.D.\ are or were affiliated with RTE, whose registry is one of the datasets audited in this
study. The audit design (direction-agnostic corrections, symmetric negative control, public-data variants
reproducible end to end) was chosen so that no conclusion depends on privileged access or favours the
data provider; the findings include under-reporting in the provider's own reference dataset. The other
authors declare no competing interests.

\section*{SUPPLEMENTAL INFORMATION INDEX}
\begin{itemize}
\item Note S1: theoretical properties of the capacity correction
\item Note S2: ground truth validity
\item Note S3: national, regional, and departmental aggregation consistency
\item Note S4: statistical robustness --- prior sensitivity and multiplicity
\item Note S5: truncation bias in the public registry
\item Note S6: imagery acquisition --- coverage and vintage effect
\item Note S7: temporal robustness
\item Note S8: the detection pipeline (DeepPVMapper)
\item Note S9: statistical lineage
\end{itemize}
\clearpage

\subsection*{Note S1: theoretical properties of the capacity correction}

\paragraph{Formal derivation.} Write $D$ for the number of raw detections, $N$ for the true number of installations, and $TP$ for the shared count of correctly detected ones, so that $P = TP/D$ and $R = TP/N$. Then
\begin{equation}
\frac{P}{R} = \frac{TP/D}{TP/N} = \frac{N}{D}.
\end{equation}
The correction factor is exactly the true-to-detected count ratio. Writing $\bar c_{\text{raw}} = C_{\text{raw}}/D$ for the mean estimated capacity across all raw detections, and $\bar c_{\text{true}} = C_{\text{true}}/N$ for the mean true capacity across all true installations,
\begin{equation}
C_{\text{adj}} = \frac{P}{R}\,C_{\text{raw}} = \frac{N}{D}\,D\,\bar c_{\text{raw}} = N\,\bar c_{\text{raw}},
\end{equation}
while $C_{\text{true}} = N\,\bar c_{\text{true}}$ by definition. Hence
\begin{equation}
C_{\text{adj}} = C_{\text{true}} \iff \bar c_{\text{raw}} = \bar c_{\text{true}}.
\end{equation}
This condition is necessary and sufficient for the correction to recover true capacity: the mean estimated capacity of raw detections must equal the mean true capacity of true installations. \textbf{H1} (detection status independent of capacity) is a clean, interpretable \emph{sufficient} condition for this equality --- if true positives, false positives, and false negatives are drawn from the same capacity distribution regardless of detection outcome, the two group means coincide automatically --- though it is formally stronger than what is strictly required. \textbf{H4} closes the remaining gap: even if \textbf{H1} holds, the true-positive component of $\bar c_{\text{raw}}$ must itself be an unbiased estimate of true capacity, which is what \textbf{H4} states.

\paragraph{Consistency of the estimator.} \textbf{H2} ensures the sampling frame for both $P$ is the department's full territory, not a sub-region; \textbf{H3} ensures the annotated samples for $R$ are drawn representatively from that frame. Under simple random sampling, the sample proportions $\hat P = TP_{\text{sample}}/n_P$ and $\hat R = TP_{\text{sample}}/n_R$ are individually unbiased for $P$ and $R$ at any sample size, and converge to them almost surely as $n_P, n_R \to \infty$ by the strong law of large numbers. Since division is continuous at $R > 0$, the continuous mapping theorem gives $\hat P/\hat R \to P/R$ almost surely, and therefore $\hat C_{\text{adj}} \to C_{\text{true}}$ under \textbf{H1}--\textbf{H4}. This proves Theorem 1 stated in Methods.

\paragraph{Finite-sample bias.} Although $\hat P$ and $\hat R$ are each unbiased, their ratio is not: for two independent random variables, $\mathbb{E}[\hat P/\hat R] = \mathbb{E}[\hat P]\cdot\mathbb{E}[1/\hat R]$, and by Jensen's inequality, $\mathbb{E}[1/\hat R] > 1/\mathbb{E}[\hat R]$, since $x \mapsto 1/x$ is convex. This is the standard finite-sample bias of a ratio estimator, of order $O(1/n_R)$, and it vanishes only as $n_R$ grows. 

The finite-sample bias of a ratio estimator is an additional reason not to report a point-corrected capacity computed from plug-in estimates of $P$ and $R$. Instead, we propagate full posterior draws of $(P,R)$ through a bootstrap (Component~2, Practical implementation), and report the resulting distribution's mean and credible interval. The distributional treatment absorbs this finite-sample bias into the reported uncertainty, rather than leaving it unquantified in a single corrected number.

\subsection*{Note S2: ground truth validity}

\paragraph{Precision.} We measure precision-label reliability using Cohen's $\kappa$\cite{cohen_coefficient_1960} between the initial labels and a blind relabelling of a stratified random national sample of 500 precision points. On this sample, raw agreement was 94.0\% and Cohen's $\kappa$ was 0.864 --- a level conventionally read as near-perfect agreement. The relabelling additionally oversampled six departments under particular scrutiny for precision (Table~S4).

Department 75 (Paris) is the exception: at $\kappa = 0.435$, its original labels were substantially less reproducible than the rest of the audit, and the corresponding shift in local precision (0.175 to 0.275 under relabelling) shows why --- its initial labels were noisy enough to be judged unreliable, and were replaced wholesale with the relabelled version. Every other department in this sample clears $\kappa \geq 0.76$, with department 50 setting the floor at 0.763 --- this is the basis for treating labels outside Paris as reliably reproducible throughout the rest of this audit.

A high $\kappa$ indicates that two annotation passes agree with each other, not that either is objectively correct. A bias shared across both passes --- for instance, a solar-thermal installation mistaken for PV in both relabelling rounds, or a reflective metal roof read as an array twice --- would be invisible to this check. This is the ceiling this test can establish without an independent, higher-fidelity ground truth; it is also the reason the possibility of residual thermal-PV confusion is treated as an acknowledged limitation elsewhere in this paper (Discussion, Limitations of the study) rather than one this check can rule out.

\subsection*{Table S4. Precision-label reproducibility for the six departments under particular
scrutiny.}
Raw agreement and Cohen's $\kappa$ between original and blind-relabelled precision annotations ($n=40$ per department); Precision (original) and Precision (relabelled) give the department's measured precision before and after relabelling. Department 75 (Paris) is the reproducibility outlier discussed in the text.

\begin{tabular}{lcccccc}
\hline
Department & $n$ & Raw agreement & Cohen's $\kappa$ & Precision (original) & Precision (relabelled) \\
\hline
75 & 40 & 0.800 & 0.435 & 0.175 & 0.275 \\
89 & 40 & 0.950 & 0.896 & 0.575 & 0.625 \\
54 & 40 & 1.000 & 1.000 & 0.625 & 0.625 \\
27 & 40 & 0.950 & 0.900 & 0.500 & 0.500 \\
50 & 40 & 0.900 & 0.763 & 0.725 & 0.675 \\
37 & 40 & 0.925 & 0.842 & 0.600 & 0.625 \\
\hline
\end{tabular}

\paragraph{Recall.} We apply the same blind re-annotation logic to the recall ground truth, on a national sample of 250 false-negative points --- ground-truth systems the pipeline failed to detect --- checked blind to the original label to identify points that do not correspond to a genuine rooftop PV system visible at the imagery vintage. Of the 250 points, 17 (6.8\%) were confirmed invalid. Unlike the precision correction, this one raises rather than lowers measured recall: an invalid ground-truth point the pipeline failed to detect was never a genuine false negative, so removing it improves, rather than worsens, the recall estimate.

The correction is applied at two levels: the 17 confirmed-invalid points are removed from the ground truth outright, and the remaining false negatives --- across all departments, not only the re-reviewed sample --- are down-weighted by $(1-\hat\tau)$, with $\hat\tau = 0.068$ the national false-negative invalidity rate measured on the 250-point sample; true positives are unaffected ($\text{weight}=1$). This propagates the sample-level finding to the full false-negative population rather than correcting only the 250 reviewed points. The correction raises national pipeline recall from 0.596 (raw) to \textbf{0.612}, corresponding the figure reported throughout this paper as pipeline recall ($\approx 0.61$; Component~1, Practical implementation). 

\paragraph{Source representativeness of the recall ground truth.} Recall ground truth is drawn from an external source (OpenStreetMap, supplemented by manual annotation) rather than sampled from the detections themselves, which raises a distinct question from label reliability: whether one source systematically performs differently from the other. At constant department, recall measured on OSM-seeded points is statistically indistinguishable from recall on manual annotations (logistic regression of detection success on source with department fixed effects: OR 0.85, $p = 0.11$); the national difference between sources is entirely compositional (see Methods, Collection of the recall samples, for why circularity with the RNI is limited). 

\subsection*{Note S3: consistency of the estimate across aggregation scales}

The correction factor is formed at the level of the reporting unit, applied to that unit's raw detected capacity, and the national total is the sum. It could also be formed higher up, on a region or on the country as a whole, by averaging the drawn unit-level $P$ and $R$ weighted by raw detected capacity. We compare the three paths.

They cannot agree exactly, and the reason is structural rather than numerical. The unit-level total is $\big(\sum_d C_d\big)\bar f$, where $\bar f$ is the weighted mean of the \emph{ratios} $P_d/R_d$; the pooled total is $\big(\sum_d C_d\big) \bar P/\bar R$, the \emph{ratio of the means}. The numerator is linear, so averaging precision before dividing costs nothing. The denominator is not, because $1/R$ is convex. It follows that if every unit in the pooled group shares the same recall, the two totals coincide identically, however dispersed precision may be. Pooling is therefore free when recall is homogeneous and costly when it is not, and to second order the relative gap is $-[\mathrm{CV}(R)^2 - \rho\,\mathrm{CV}(P)\mathrm{CV}(R)]$ ($\mathrm{CV}$ denotes the coefficient of variation and $\rho$ the correlation between precision and
recall, both taken across the sub-units being merged and weighted by their raw detected capacity).

Empirically the three levels agree well within sampling noise. Forming the factor at the regional level lowers the national total by 1.31\%, and at the national level by 1.26\%, against a 95\% credible interval whose half-width is 1.8\% at the unit level (Figure~S4). The gap is nonetheless signed rather than random: pooling always yields a smaller factor. Its magnitude follows the mechanism. Nationally $\mathrm{CV}(P) = 0.149$, $\mathrm{CV}(R) = 0.143$ and $\rho(P,R) = +0.55$, for which the second-order expression predicts $-0.9\%$, the right sign and order. Across the 13 regions the pooling gap correlates with the within-region dispersion of \emph{recall} (Spearman $\rho = +0.86$) and not with that of precision ($\rho = -0.08$).

Two remarks follow. First, the wording matters: recall itself is not underestimated by pooling, since $\bar R$ is an honest weighted mean. What is understated is $1/\bar R$ as an estimate of $\overline{1/R}$, hence the correction factor and hence the corrected capacity. Second, there is no trade-off to arbitrate. Forming the factor at the finest level for which a calibration exists is weakly better in all cases, and the variance argument that would favour pooling is already handled by the empirical-Bayes prior, which shrinks small-sample units on $P$ and $R$ separately without forcing the ratio to be pooled.

The unit level reproduces the published national total to machine precision, which makes this note a non-regression test of the recalibration pipeline. Under a Jeffreys prior every total shifts up by 1.9\% while the ranking and sign of the between-level gaps are unchanged, so the choice of a mildly informative prior does not drive the result; it is, if anything, the more conservative of the two with respect to under-reporting.

\noindent\includegraphics[width=0.95\textwidth]{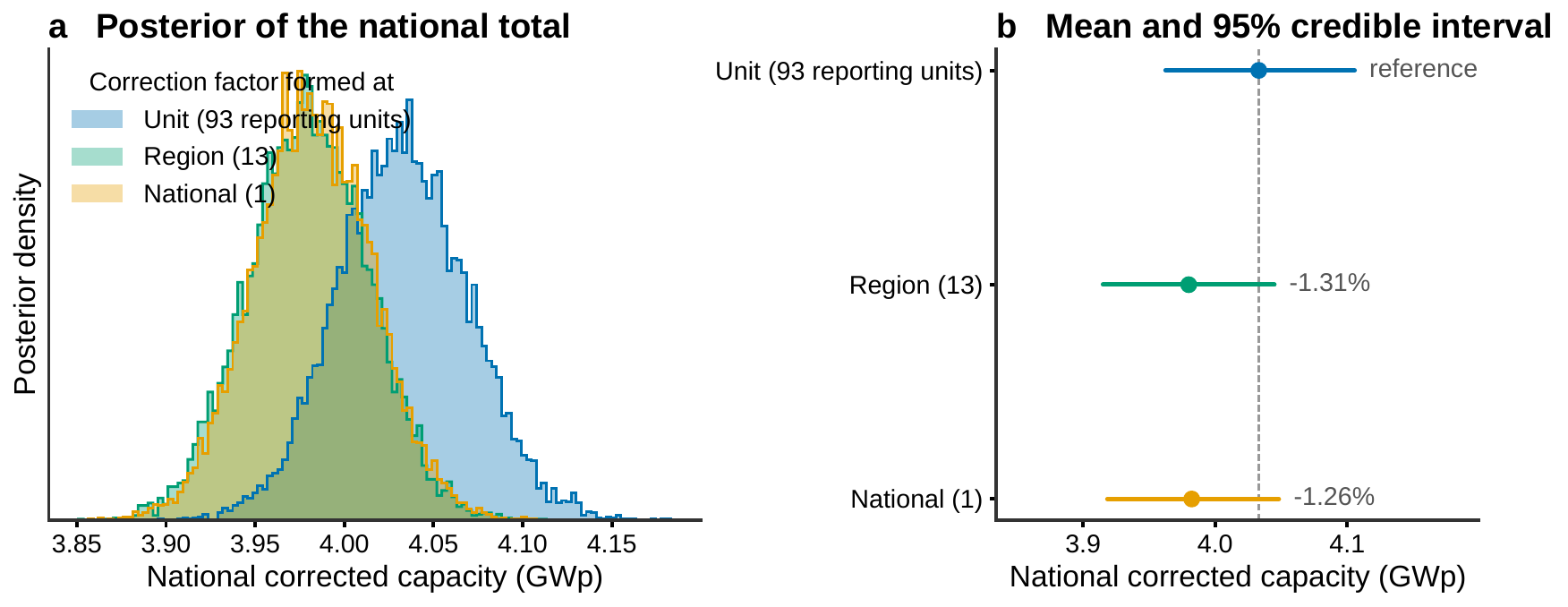}
\subsection*{Figure S8. National corrected capacity under three aggregation scales.}
(a) Posterior distribution of the national corrected capacity when the correction factor is formed at the level of the reporting unit (blue, 93 units), of the region (green, 13) or of the country (orange, 1); 10,000 bootstrap draws each, empirical-Bayes prior. (b) Means and 95\% credible intervals of the same three distributions, with the published total as a dashed line and the relative gap to the unit-level reference annotated. The two pooled levels sit 1.3\% below the reference, roughly two thirds of the credible half-width, and are indistinguishable from one another.

\subsection*{Note S4: statistical robustness}

\paragraph{Prior sensitivity.} All results were recomputed under the unpooled Jeffreys prior $\mathrm{Beta}(0.5, 0.5)$ in place of the empirical-Bayes prior: shrinkage only affects small-sample units, and every hard-core unit remains flagged. Pooling stabilises the measurement without being able to manufacture or attenuate the audit signal.

\paragraph{Multiplicity.} At a two-sided 1\% flagging threshold on 93 units, the expected number of false flags under the global null is below one. The observed configuration (25 below / 8 above at the baseline specification; 18/7 after unanimity) cannot be produced by multiplicity; the symmetric negative control, subjected to the identical criteria, provides the empirical false-discovery bound and remained stable (7 units) across every revision of the reference data during the study.

\subsection*{Note S5: truncation bias in the public registry}

The public registry applies a confidentiality rule: a municipality with fewer than ten installations does not appear in the published municipal view. Its capacity is not missing from the registry, only from the public view at that resolution, and it remains counted in the departmental totals. The same source therefore yields two references. Summing the published municipalities, as a portal user would, gives the truncated view; using the published departmental totals gives the corrected one. The audit is identical in both cases, same corrected estimate, same battery, same thresholds, so the difference between them measures an artefact of the reference rather than of the method.

The rule removes 14,019 municipalities, 40.3\% of the total, carrying 363~MWp of detected capacity or 11.6\% of the national detected fleet. The affected municipalities are the smallest ones, with a median population of 211 against 873 elsewhere, which concentrates the censoring in rural areas (Figure~S9a). The censored share of detected capacity has a median of 13\% across reporting units, exceeds 40\% in eleven of them and reaches 100\% in Corse-du-Sud.

The consequence for a local audit is large and runs in one direction only. Auditing the truncated municipal view flags 32 units as under-reporting; the same audit against the departmental totals flags 8. Twenty-four units, three quarters of those flagged on the public view, lose their flag once the censoring is undone, and no unit acquires one. The eight that survive, Corr\`eze, Creuse, Corse-du-Sud, Haute-Corse, Loz\`ere, Haute-Sa\^one, Paris with its inner ring, and Territoire de Belfort, are the only ones whose diagnosis implicates the registry rather than its published view. This note reports under-reporting alone; correcting the truncation also produces over-reported units, but those arise from the annual dating of the departmental totals, a distinct artefact.

The eight units that survive the correction are not a new finding but a corroboration. All eight, Corr\`eze, Creuse, Corse-du-Sud, Haute-Corse, Loz\`ere, Haute-Sa\^one, Paris with its inner ring and Territoire de Belfort, belong to the hard core established independently against the grid connection data of the transmission operator. Two registries, compiled by different bodies through different reporting chains, audited separately against the same corrected estimate, and the public registry's under-reported set falls entirely inside the connection data's. Nothing in the design forces this agreement, and it is the strongest evidence that the local diagnosis tracks the fleet rather than the idiosyncrasies of one reference.

\noindent\includegraphics[width=0.95\textwidth]{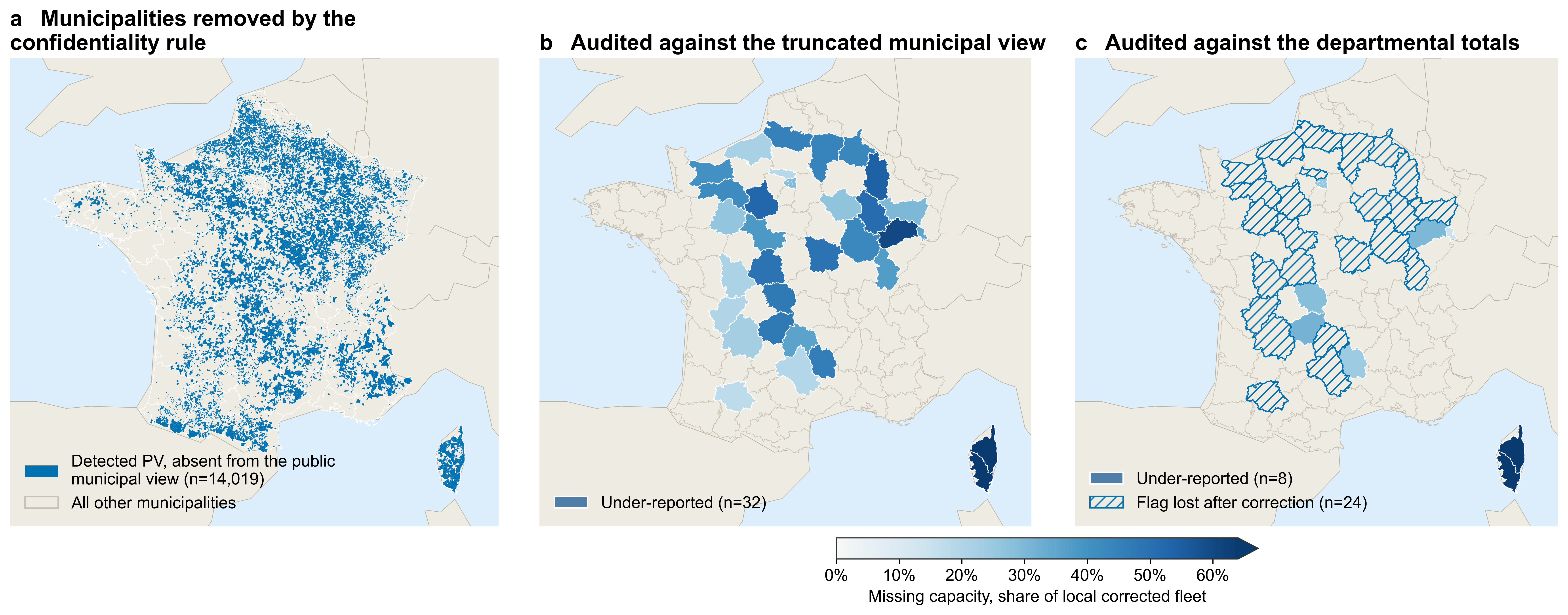}
\subsection*{Figure S9. The public registry's confidentiality rule and what it does to a local audit.}
(a) Municipalities removed by the confidentiality rule: detected PV capacity located in the 14,019 municipalities absent from the public municipal view, against all other municipalities. (b) Reporting units flagged as under-reporting when the audit is run against the truncated municipal view, shaded by the missing capacity as a share of the local corrected fleet. (c) The same audit against the departmental totals, on a common colour scale capped at the 90th percentile so that Corse-du-Sud, censored in full, does not compress the remaining contrast; hatching marks the 24 units whose flag disappears once the censoring is undone. Over-reported units are omitted from panels (b) and (c).

\subsection*{Note S6: imagery acquisition --- coverage and vintage effect}

\paragraph{No vintage effect.} Pipeline recall shows no dependence on imagery acquisition year once local detection quality is controlled: a likelihood-ratio test of the year effect on top of departmental F1 yields $p = 0.54$; the raw recall gradient across the 2022--2025 campaigns is geographic composition. The model generalises across acquisition campaigns, a prerequisite for comparing units imaged in different years.

The absence of distribution shift may be due to the fact that the radiometric characteristics of IGN's orthoimagery remained stable over the recent years: BDAPPV\cite{kasmi_crowdsourced_2023} features IGN orthoimagery from annotations campaigns slightly older than the 2022--2025 study window. The result above rules out one specific way that variation could bias the audit --- a department's flag status tracking its imagery's age rather than a genuine registry gap.

\noindent\includegraphics[width=0.65\textwidth]{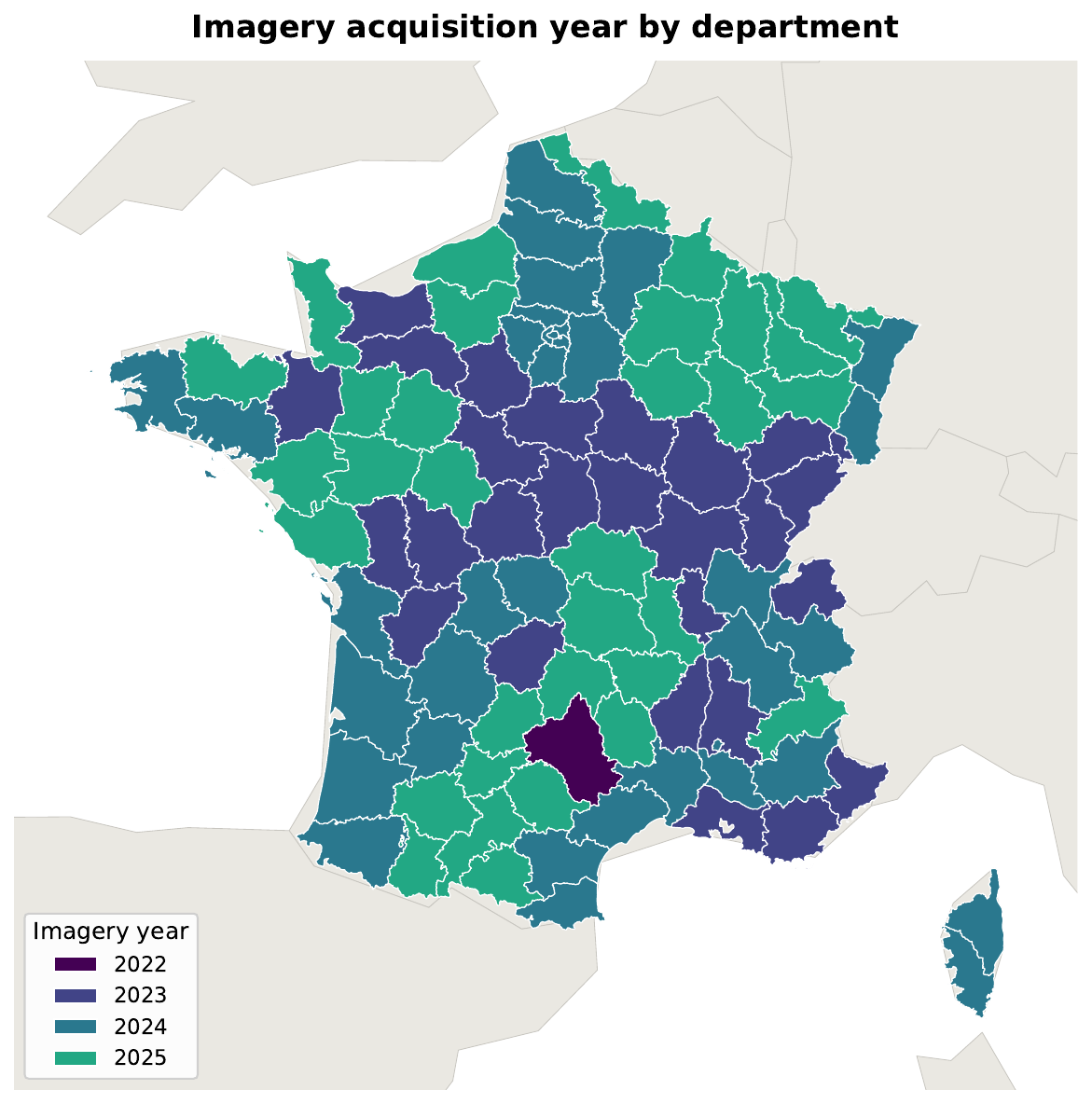}\\
\noindent\includegraphics[width=0.75\textwidth]{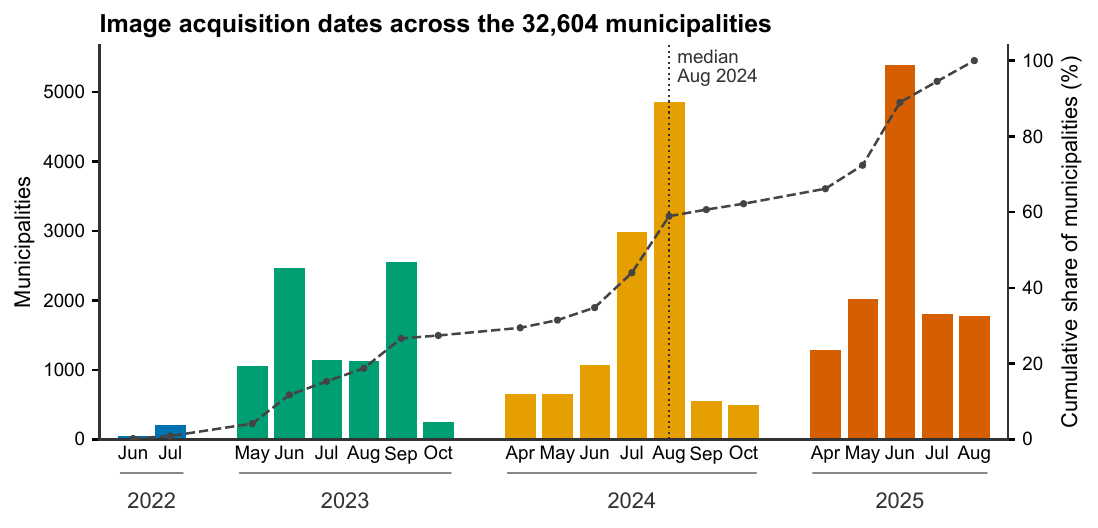}
\subsection*{Figure S10. Coverage and vintage of the aerial orthoimagery used in this study.} Top: the acquisition year of the most recent available imagery tile, by department. Bottom: distribution
of the month of acquisition across municipalities, illustrating the concentration of flight campaigns in the summer months described in Methods (Temporal alignment: implementation).

\subsection*{Note S7: temporal robustness}

The audit's day-level temporal alignment (Methods, Component~3) is stress-tested two ways: first, by checking that the flagged gaps survive plausible bounds on connection lag in both directions; second, by checking that the audit's conclusions survive degraded date granularity, down to the month-end cutoffs available in most other registries.

\subsubsection*{Connection lags}

\paragraph{Connection lag, downstream (under-reporting side).} For every hard-core unit, the flagged gap is compared to the transit mass: registry capacity connected between the local imagery date and the registry extraction (13 October 2025). Under the maximally generous assumption that this entire mass was already installed at the imagery date, the gap survives in 12 of 18 units; restricting to a realistic three-month connection window, it survives in 18 of 18 --- the transit at three months (0.1--3~MWp per unit) is everywhere an order of magnitude below the gaps (2--36~MWp). The single borderline case is Bouches-du-Rh\^one (gap 12.4 vs three-month transit 12.1~MWp), which is also conversion-sensitive and flagged as fragile on both axes in the tagging. Cantal, re-imaged in August 2025, retained its full gap over two consecutive temporal alignments.

\paragraph{Connection lag, upstream (over-reporting side).} The mirror artefact --- administrative connection dates preceding physical installation --- is bounded symmetrically: for every negative-control unit, the registry excess exceeds all capacity connected in the three months \emph{preceding} the image (7/7; e.g.\ Haute-Garonne: excess 27.2 vs pre-image mass 5.8~MWp). The temporal explanation is closed in both directions.

\paragraph{RNI annual attribution is a level artefact.} The truncation-corrected RNI, dated 31 December, overshoots the mid-year corrected estimate by 10.8\% nationally (Note~S5). Registry stock growth is quasi-uniform across departments ($\sim$25\%~yr$^{-1}$), and the year-over-year growth rate does not predict which units flag as over-reported (Pearson $r = +0.02$, $p = 0.84$; above-flagged units' median growth 25.0\% vs 24.9\% elsewhere, Mann-Whitney $p = 0.84$): the annual dating convention shifts the level of the comparison, not its cross-section.

\subsubsection*{Sensitivity to dating}

\paragraph{Sensitivity to month-end registry dating.} The French demonstration attains the gold standard of temporal alignment: day-level connection dates, allowing the registry series to be stopped at the exact acquisition date of each municipality. Other registries may publish, at best, month-end stocks. This paragraph quantifies what the audit loses when only a month-end cutoff (31/MM) is available, by re-aggregating the TSO registry to month-end cutoffs offset by $-3$ to $+3$ months around the month of each unit's acquisition date (offset 0 being what a monthly registry would naturally provide), and re-evaluating every unit's status against its unchanged calibration posterior.

The audit's local conclusions are robust to month-level dating. At the natural month-end cutoff (offset 0), 16 of the 18 hard-core units remain flagged; across the entire $\pm$3-month sweep, survival ranges from 18/18 (reference minimised) to 14/18 (reference maximised, the case most adverse to under-reporting flags). The negative control is invariant: 7/7 units at every one of the seven offsets. This result shows that  the audit does not manufacture flags in either direction under dating stress. Unit statuses overall remain unchanged for the large majority of the 93 units at every offset. This sweep, incidentally, also speaks to a distinct concern raised by the fragmentation mechanism (Lessons from the French case study): because some ELDs report to RTE only twice a year rather than monthly, a natural alternative account of the ELD gap is a dating artefact, whereby capacity connected between two biannual filing dates would appear temporarily absent from the registry. The $\pm$3-month cutoff sweep already brackets this lag, and the hard core survives it in full: the ELD gap therefore cannot be a reporting-cadence artefact, reinforcing the integration-based account favoured in the main text over a pure dating explanation.

The national-level statement, by contrast, is dating-sensitive: the national surplus swings from $+8.5\%$ (cutoff 3 months early) to $-3.1\%$ (3 months late) across the sweep --- a further reason why national figures are reported here as scale-agreement with brackets rather than as claims, and why fine-grained comparisons against annually-dated registries are dominated by the dating artefact.

The senstivity appears to be predictable, making the result exportable. The relative width of the dating bracket per unit --- (reference$_{+3}$ $-$ reference$_{-3}$)/reference$_0$, median 11.5\%, maximum 37.4\% --- is mechanically the capacity connected during the window, and correlates at $r = 0.998$ with local registry growth. 

A rule of thumb for a prospective audit can be stated as follows: the reference uncertainty induced by dating granularity equals the local growth rate accumulated over that granularity. At French growth rates ($\sim$25\% yr$^{-1}$), month-level dating costs $\sim$2\% of reference per month of uncertainty --- absorbed by the audit's 99\% credible thresholds --- whereas in a market growing severalfold faster, the same granularity could dominate the signal: the condition is checkable from public data before any annotation effort is spent.

\noindent\includegraphics[width=0.95\textwidth]{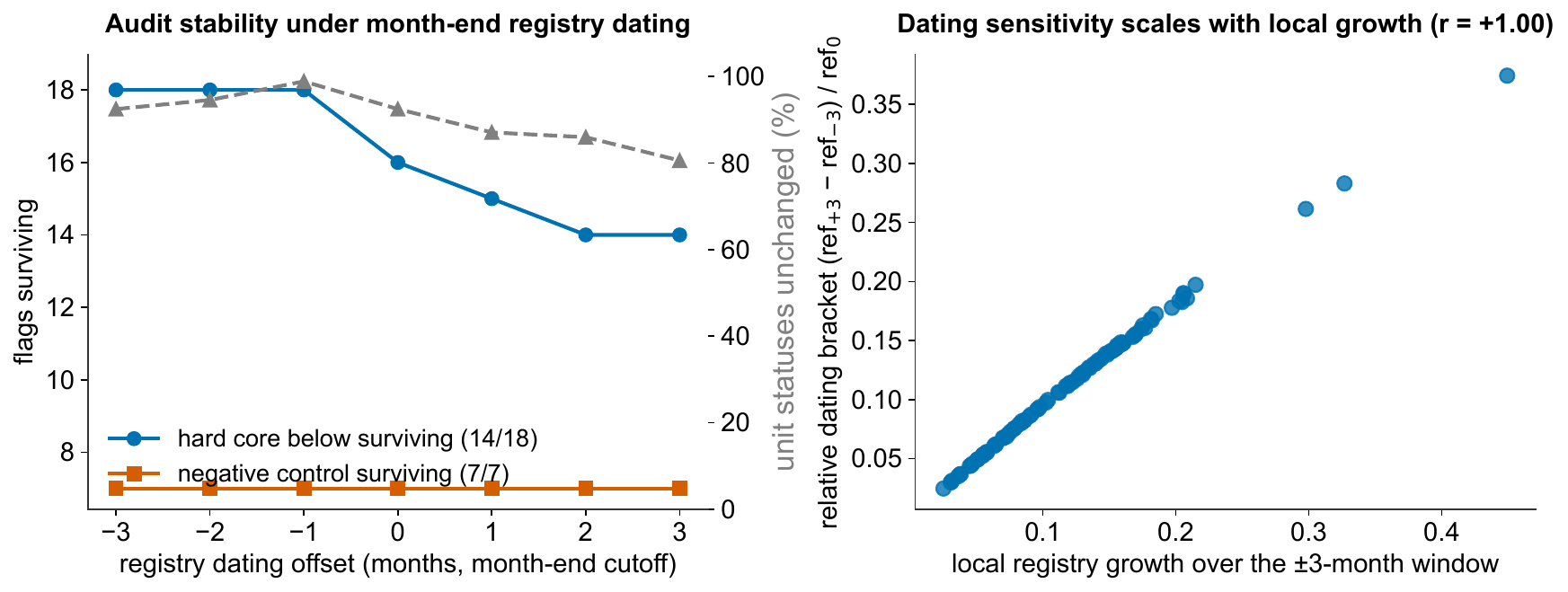}
\subsection*{Figure S11. Audit stability under month-end registry dating.}
Left: hard-core and negative-control survival, and share of unchanged unit statuses, as the month-end cutoff is offset by $-3$ to $+3$ months around the acquisition date. Right: the relative dating bracket per unit against local registry growth over the window ($r = 0.998$): dating sensitivity is the accumulated local growth, a quantity checkable in advance by any prospective audit.

\subsection*{Note S8: the detection pipeline (DeepPVMapper)}

This note summarises the detection pipeline underlying the audit; full details are published in refs.~\cite{kasmi_towards_2022,kasmi_deeppvmapper_2023} and the training data in ref.~\cite{kasmi_crowdsourced_2023}. The pipeline (Figure~S12) takes as input the IGN 20-cm orthoimagery (BD ORTHO) and building footprints (BD TOPO), and outputs a georeferenced registry of PV arrays with location, projected surface, tilt, azimuth and estimated capacity.

\textbf{Classification and segmentation.} Detection follows the two-stage design introduced by DeepSolar~\cite{yu_deepsolar_2018}: image tiles are first classified for the presence of PV, and positive tiles are passed to a semantic segmentation model whose masks are vectorised into polygons. Both models are fine-tuned on BDAPPV, a crowdsourced dataset of French rooftop installations~\cite{kasmi_crowdsourced_2023}, initialised from the weights of ref.~\cite{mayer_deepsolar_2020}. The classifier is an Inception-v3 and the segmenter a DeepLab-v3~\cite{kasmi_towards_2022}. On the held-out test set, the pipeline reaches a classification F1 of 0.84 and a segmentation IoU of 0.86, competitive with published systems operating at finer ground sampling distances (Table~S5)~\cite{mayer_3d-pv-locator_2022,malof_mapping_2019,zech_predicting_2020,parhar_hyperionsolarnet_2021}. The audit of the main text deliberately does not rely on these test-set figures: as established in refs.~\cite{kasmi_towards_2022,kasmi_enhancing_2024}, test-set accuracy substantially overestimates accuracy over the mapping area, which is why the audit protocol measures pipeline-level precision and recall directly (Component~1).

\textbf{Characteristics extraction.} From each polygon, the pipeline extracts the location and projected surface directly, and infers the tilt angle from a look-up table (LUT) computed on the BDPV metadata database~\cite{kasmi_crowdsourced_2023}: France is divided into geographic cells, and for each cell and each of four projected-surface clusters, the LUT stores the mean tilt observed in the metadata --- capturing both the north--south steepening of tilts and the fact that smaller arrays are steeper (Figure~S13)~\cite{kasmi_towards_2022}. Azimuth and installed capacity are then derived from the polygon geometry and the tilt, the capacity through the surface-to-capacity relation whose coefficient is discussed in the main text; the extraction stage follows the methods consolidated in PyPVRoof~\cite{tremenbert_pypvroof_2023}. Post-processing removes detections not attached to a building (BD TOPO) and applies the perimeter thresholds; in the deployed configuration used here, the product's effective detection floor is $\approx$11 m$^2$ and the upper perimeter bound 36 kWp.

\subsection*{Table S5. Test-set detection metrics of published rooftop-PV mapping systems}
\begin{tabular}{lccc}
\hline
Work & Classification (F1) & Segmentation (IoU) & GSD (cm/pixel) \\
\hline
3D-PV-Locator~\cite{mayer_3d-pv-locator_2022} & 0.87 & 0.74 & 10 \\
Malof et al.~\cite{malof_mapping_2019} & --- & 0.67 & 30 \\
Zech \& Ranalli~\cite{zech_predicting_2020} & 0.82 & --- & 10 \\
HyperionSolarNet~\cite{parhar_hyperionsolarnet_2021} & 0.97 & 0.86 & 10 \\
DeepPVMapper (this work)~\cite{kasmi_deeppvmapper_2023} & 0.84 & 0.86 & 20 \\
\hline
\end{tabular}
\bigskip\\
\noindent
GSD denotes the ground sampling distance of the input imagery used by the different works. Source:~\cite{kasmi_towards_2022}.

\noindent\includegraphics[width=0.85\textwidth]{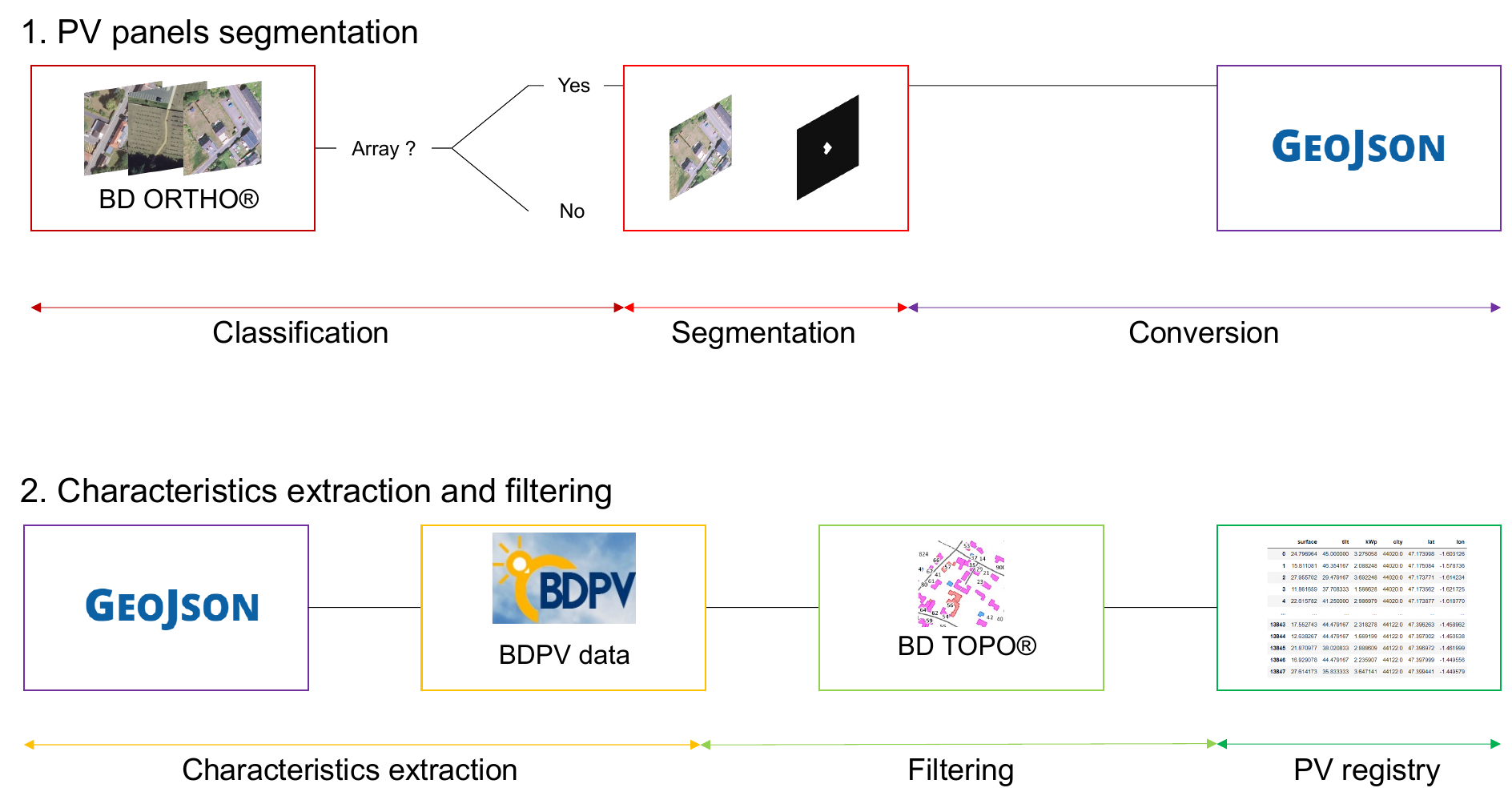}
\subsection*{Figure S12. The DeepPVMapper pipeline.}
Classification and segmentation fine-tuned on crowdsourced training data, characteristics extraction (location, surface, tilt via look-up table, azimuth, capacity), and post-processing against building footprints and perimeter thresholds. Source:~\cite{kasmi_towards_2022}.

\noindent\includegraphics[width=0.75\textwidth]{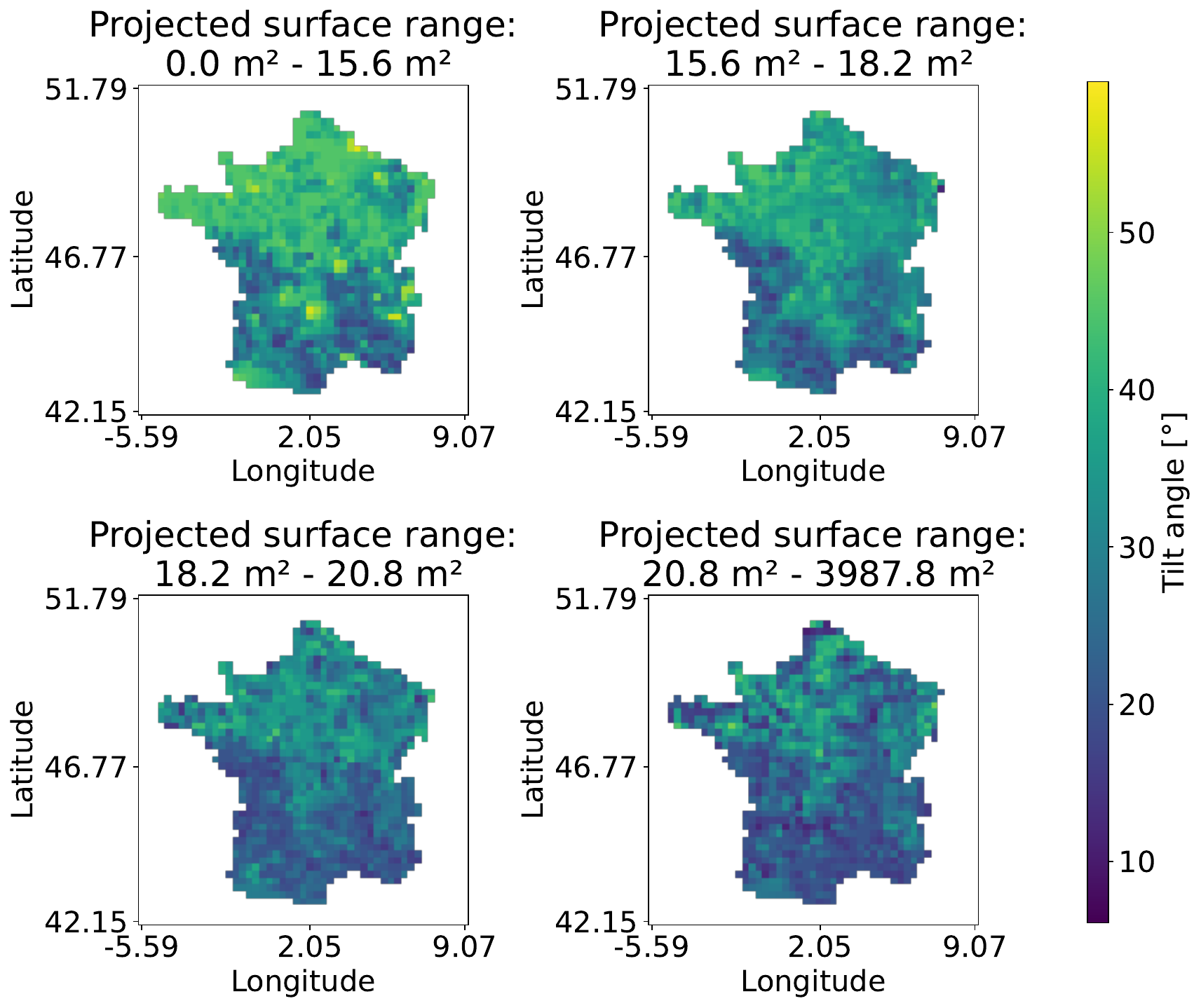}
\subsection*{Figure S13. Tilt look-up table.}
Mean tilt per geographic cell, stratified by projected-surface cluster --- tilts steepen northward and for smaller arrays. Source:~\cite{kasmi_towards_2022}.

\subsection*{Note S9: statistical lineage}

The correction methodology introduced in this work draws on three established statistical traditions, which are introduced in this section.

\paragraph{Dual-system and capture--recapture estimation.} A long-standing problem in demography and epidemiology is estimating the true size of a population from incomplete enumerations of it. The classical solution, dual-system estimation, combines two independent, imperfect counts of the same population: where they overlap, and where they diverge, is informative about how many individuals neither count captured\cite{hook_regal_1992,hook_regal_1995}. Capture--recapture methods extend this logic to more general settings, using the rate of re-observation across repeated samples to infer an unobserved total\cite{chao_2001}. The idea that two imperfect enumerations, compared to one another, can reveal more than either alone motivates the overall structure of this audit, comparing a corrected detection product against a registry.

\paragraph{Detection-probability modelling.} A related problem in ecological surveys is that failing to observe a species at a site does not mean it is absent: it may simply have gone undetected. Occupancy modelling separates these two questions explicitly, estimating a detection probability alongside the quantity of interest, rather than treating a non-detection as evidence of absence\cite{mackenzie_estimating_2002}. Survey-design methodology in this tradition also addresses how such studies should be structured and sampled at scale\cite{pollock_large_2002}, and hierarchical Bayesian treatments formalise how detection probability and the underlying quantity are estimated jointly, with uncertainty carried through both\cite{royle_hierarchical_2008}. This distinction between non-detection and absence is the logic behind treating every undetected installation as a correctable false negative, rather than as evidence that no installation exists (Component~1--2).

\paragraph{Accuracy-adjusted area estimation.} In remote sensing of land cover, a classified map is rarely accurate as published: some pixels are misclassified, and simply trusting the map's own tally overstates or understates the true area of a given class. The standard remedy draws a stratified reference sample, builds a confusion matrix from it, and uses that matrix to correct the map's area estimate, with the correction's own uncertainty reported alongside it\cite{card_1982,olofsson_making_2013,olofsson_good_2014}. This is the closest structural match to the correction applied throughout this paper: a confusion matrix, measured on a validated sample, corrects a raw estimate, and the correction carries a propagated interval rather than a single number (Component~1--2).

\clearpage
\bibliography{references}
\end{document}